\documentclass[12pt]{article}
\usepackage{a4wide}
\usepackage{authblk}
\usepackage{graphicx}
\usepackage{amsmath}
\usepackage{bbm}
\usepackage[usenames]{color}
\usepackage{hyperref}

\title{\textbf{Predictions on the second-class current decays $\tau^{-}\to\pi^{-}\eta^{(\prime)}\nu_{\tau}$}}
\author[a,b]{R. Escribano\thanks{rescriba@ifae.es}}
\author[b]{S. Gonz\`{a}lez-Sol\'{i}s\thanks{sgonzalez@ifae.cat}}
\author[c]{P. Roig\thanks{proig@fis.cinvestav.mx}}
\affil[a]{{\normalsize\textit{Grup de F\'{\i}sica Te\`orica, Departament de F\'{\i}sica, 
Universitat Aut\`onoma de Barcelona, E-08193 Bellaterra (Barcelona), Spain}}}
\affil[b]{{\normalsize\textit{Institut de F\'{\i}sica d'Altes Energies (IFAE), 
The Barcelona Institute of Science and Technology, Campus UAB, 08193 Bellaterra (Barcelona), Spain}}}
\affil[c]{{\normalsize\textit{Departamento de F\'{i}sica, 
Centro de Investigaci\'{o}n y de Estudios Avanzados del Instituto Polit\'{e}cnico Nacional, 
Apartado Postal 14-740, 0700 M\'{e}xico D.F., M\'{e}xico}}}

\def\be{\begin{equation}}
\def\ee{\end{equation}}
\def\bea{\begin{eqnarray}}
\def\eea{\end{eqnarray}}

\begin{document}
\maketitle

\abstract{
We analyze the second-class current decays $\tau^{-}\to\pi^{-}\eta^{(\prime)}\nu_{\tau}$
in the framework of Chiral Perturbation Theory with resonances.
Taking into account $\pi^{0}$-$\eta$-$\eta^{\prime}$ mixing,
the $\pi^{-}\eta^{(\prime)}$ vector form factor is extracted, in a model-independent way,
using existing data on the $\pi^{-}\pi^{0}$ one.
For the participant scalar form factor,
we have considered different parameterizations ordered according to their increasing fulfillment of
analyticity and unitarity constraints.
We start with a Breit-Wigner parameterization dominated by the $a_{0}(980)$ scalar resonance
and after we include its excited state, the $a_{0}(1450)$.
We follow by an elastic dispersion relation representation through the Omn\`{e}s integral.
Then, we illustrate a method to derive a closed-form expression for the
$\pi^{-}\eta$, $\pi^{-}\eta^{\prime}$ (and $K^{-}K^{0}$) scalar form factors in a coupled-channels treatment.
Finally, predictions for the branching ratios and spectra are discussed emphasizing the error analysis.
An interesting result of this study is that both $\tau^{-}\to\pi^{-}\eta^{(\prime)}\nu_{\tau}$ decay channels
are promising for the soon discovery of second-class currents at Belle-II.
We also predict the relevant observables for the partner $\eta^{(\prime)}_{\ell 3}$ decays,
which are extremely suppressed in the Standard Model.}

\tableofcontents

\section{Introduction}
According to Weinberg \cite{Weinberg:1958ut},
non-strange weak $(V-A)$ hadronic currents can be divided into two types depending on their $ G$-parity:
i) first class currents, with the quantum numbers $J^{PG}=0^{++},0^{--},1^{+-},1^{-+}$;
ii) second class currents (SCC), which have $J^{PG}=0^{+-},0^{-+},1^{++},1^{--}$.
The former completely dominate weak interactions since there has been no evidence of the later in Nature so far.

In the Standard Model (SM) SCC come up with an isospin-violating term which heavily suppresses the interaction
and the eventual sensitivity to new physics
({\textit{i.e.}}~by a charged Higgs contribution to the $\pi\eta^{(\prime)}$ scalar form factors) may be enhanced.

One tentative scenario to look for such kind of currents is through the rare hadronic decays of the $\tau$ lepton
$\tau^{-}\to\pi^{-}\eta\nu_{\tau}$ and $\tau^{-}\to\pi^{-}\eta^{\prime}\nu_{\tau}$ \cite{Leroy:1977pq}
for which some experimental upper bounds already exist.
For the $\pi^{-}\eta$ decay mode,
BaBar, Belle and CLEO collaborations have reported the branching ratio upper limits of
$9.9\cdot10^{-5}$ at $95\%$ CL \cite{delAmoSanchez:2010pc}, $7.3\cdot10^{-5}$ at $90\%$ CL \cite{Hayasaka:2009zz}
and of $1.4\cdot10^{-4}$ at $95\%$ CL \cite{Bartelt:1996iv}, respectively.
Actually, $\tau^{-}\to\pi^{-}\eta\nu_{\tau}$ belongs to the discovery modes list of the near future
super-B factory Belle II \cite{Belle2} for which we advocate the measurement.
Regarding the $\pi^{-}\eta^{\prime}$ channel,
BaBar obtained a new upper bound, $4.0\cdot10^{-6}$ at $90\%$ CL \cite{Lees:2012ks},
that slightly improved its previous value $7.2\cdot10^{-6}$ at $90\%$ CL \cite{Aubert:2008nj}.
Also CLEO quoted the upper limit $7.4\cdot10^{-5 }$ at $90\%$ CL \cite{Bergfeld:1997zt} in the nineties.
Historically, $\tau^{-}\to\pi^{-}\eta\nu_{\tau}$ decays attracted a lot of attention at the end of the eighties
when existing measurements hinted at abnormally large branching fractions into final states containing $\eta$ mesons,
and a preliminary announcement by the HRS Coll.~advocated for an $\mathcal{O}(\%)$ decay rate
into the $\pi^{-}\eta$ decay mode, which was against theoretical expectations \cite{Pich:1987qq}.
Later on, the situation settled \cite{Pich:2013lsa} and these decays remained undiscovered even at the first generation
B-factories BaBar and Belle,
where the background from other competing modes such as $\tau^{-}\to\pi^{-}\pi^0\eta\nu_{\tau}$
\cite{Inami:2008ar, Dumm:2012vb} veiled the SCC signal.
According to our results, their discovery (through either of the $\tau^{-}\to\pi^{-}\eta^{(\prime)}\nu_{\tau}$ decay channels)
should be finally possible at Belle-II,
thanks to the fifty times increased luminosity of Belle-II \cite{Abe:2010gxa} with respect to its predecessor.
The implementation of theory predictions for these modes in the TAUOLA version used by the Belle
\cite{Shekhovtsova:2012ra} Collaboration will help to accomplish this task.

From the theoretical perspective, the spin-parity of the $\pi^{-}\eta^{(\prime)}$ system,
$J^{P}$, is $0^{+}$ or $1^{-}$ depending whether the system is in $S$- or $P$-wave, respectively.
However, the $G$-parity of the system is $-1$,  which is opposed to the vector current that drives the decay in the SM.
Therefore, the $S(P)$-wave of the $\pi^{-}\eta^{(\prime)}$ system gives $J^{PG}=0^{+-}(1^{--})$,
which can only be realized through a SCC independently of possible intermediate resonant states.
Previous theoretical analysis estimated the branching ratio to be of the order of $10^{-5}$
and within the range $10^{-8}$ to $10^{-6}$ for the $\pi^{-}\eta$ and $\pi^{-}\eta^{\prime}$ modes, respectively.
In this work, we revisit these processes benefited from our previous experiences in describing dimeson $\tau$ decays data
\cite{Boito:2008fq,Boito:2010me,Dumm:2013zh,Escribano:2013bca,Escribano:2014pya,Escribano:2014joa,
Gonzalez-Solis:2015nfa}.
Here, the main subject of our study is the theoretical construction of the participant vector and scalar form factors.
Our initial approach is carried out within the framework of the Chiral Perturbation Theory (ChPT) \cite{ChPT}
including resonances (RChT) \cite{Ecker:1988te}.
On a second stage, we take advantage of the global analysis of the $U(3)\otimes U(3)$ one-loop meson-meson scattering
in the frame of RChT performed in Ref.~\cite{Guo:2011pa} to calculate the scalar form factors from dispersion relations
based on arguments of unitarity and analyticity.
In particular, we will first take into account elastic final state interactions through the Omn\`{e}s solution
\cite{Omnes:1958hv} for describing the $\pi^{-}\eta$ and $\pi^{-}\eta^{\prime}$ scalar form factors (SFF), respectively.
Then, we consider the effect of coupled channels in the former system for studying inelasticities.
Afterwards, we will also consider the $K^{-}K^{0}$ threshold, whose coupling to the intermediate scalar resonance
is presumably large \cite{Guo:2011pa}, and couple it to both $\pi^{-}\eta$ and $\pi^{-}\eta^{\prime}$ SFFs independently.
Finally, the three coupled-channels case will we addressed.
Several ways of solving coupled channels form factors have been considered in literature;
some use iterative methods \cite{Donoghue:1990xh,Moussallam:1999aq,Jamin:2001zq,Celis:2013xja},
while others employ closed algebraic expressions
\cite{Iwamura:1976fc,Babelon:1976kv,Iwamura:1977ds,Kamal:1979be,Kamal:1980mw,Kamal:1987nm,
Kamal:1988ub,Oller:2000ug}.
The second alternative will be followed in this work.
See also Ref.~\cite{Albaladejo:2015aca} for a recent description based on dispersive techniques.

The paper is organized as follows.
In Section \ref{hadronicmatrix},
we define the hadronic matrix element in terms of the vector and scalar form factors
and give the expression for the differential decay width. In Section \ref{VectorFormFactor},
we derive the $\pi^{-}\eta^{(\prime)}$ vector form factor (VFF) within RChT
by considering mixing within the $\pi^{0}$-$\eta$-$\eta^{\prime}$ system.
In our approach, the VFFs appear to be an isospin-violating factor times the $\pi^{-}\pi^{0}$ form factor
for which we will employ its experimental determination arising from the well-known first-class current
$\tau^{-}\to\pi^{-}\pi^{0}\nu_{\tau}$ decay.
We devote Section \ref{ScalarFormFactors} to the computation of the corresponding scalar form factors.
We start with a simple Breit-Wigner parameterization and then consider a dispersion relation obeying unitarity,
first in the elastic single channel case through the Omn\`{e}s solution and then taking into account coupled-channel effects.
The spectra and predictions for the branching ratios are given in Section \ref{predictions}.
Also in this section, we will briefly discuss the crossing symmetric $\eta^{(\prime)}_{\ell 3}$ decays,
$\eta^{(\prime)}\to\pi^{+}\ell^{-}\bar{\nu}_{\ell}$ $(\ell=e,\mu)$, for which branching ratio predictions will be given as well.
Finally, we present our conclusions in Section \ref{conclusions}.

\section{Hadronic matrix element and decay width}
\label{hadronicmatrix}

The amplitude of the decay $\tau^{-}\to\pi^{-}\eta^{(\prime)}\nu_{\tau}$
in terms of the hadronic matrix element reads
\be
\mathcal{M}=\frac{G_{F}}{\sqrt{2}}V_{ud}\bar{u}(p_{\nu_{\tau}})\gamma_{\mu}
(1-\gamma_{5})u(p_{\tau})\langle\pi^{-}\eta^{(\prime)}|\bar{d}\gamma^{\mu}u|0\rangle\ ,
\ee
where the $\pi^{-}\eta^{(\prime)}$ matrix element of the vector current follows the convention of Ref.~\cite{Gasser:1984ux},
\be
\langle\pi^{-}\eta^{(\prime)}|\bar{d}\gamma^{\mu}u|0\rangle=
c^{V}_{\pi^{-}\eta^{(\prime)}}\left[(p_{\eta^{(\prime)}}-p_{\pi^{-}})^{\mu}F_{+}^{\pi^{-}\eta^{(\prime)}}(s)-
(p_{\eta^{(\prime)}}+p_{\pi^{-}})^{\mu}F_{-}^{\pi^{-}\eta^{(\prime)}}(s)\right]\ ,
\label{matrixelement}
\ee
with $c^{V}_{\pi^{-}\eta^{(\prime)}}=\sqrt{2}$, $s=q^{2}=(p_{\eta^{(\prime)}}+p_{\pi^{-}})^{2}$
and $F_{+(-)}^{\pi^{-}\eta^{(\prime)}}(s)$ the two Lorentz-invariant vector form factors.
However, instead of $F_{-}^{\pi^{-}\eta^{(\prime)}}(s)$,
the scalar form factor $F_{0}^{\pi^{-}\eta^{(\prime)}}(s)$ is usually employed,
which arises as a consequence of the non-conservation of the vector current.
That is, taking the divergence on the left-hand side of Eq.~(\ref{matrixelement})
we get
\be
\langle\pi^{-}\eta^{(\prime)}|\partial_{\mu}(\bar{d}\gamma^{\mu}u)|0\rangle=
i(m_{d}-m_{u})\langle\pi^{-}\eta^{(\prime)}|\bar{d}u|0\rangle\equiv
i\Delta_{K^{0}K^{+}}^{\rm QCD}c_{\pi^{-}\eta^{(\prime)}}^{S}F_{0}^{\pi^{-}\eta^{(\prime)}}(s)\ ,
\label{divergence1}
\ee
with $c_{\pi^{-}\eta}^{S}=\sqrt{2/3}$, $c_{\pi^{-}\eta^{\prime}}^{S}=2/\sqrt{3}$ and $\Delta_{PQ}=m_{P}^{2}-m_{Q}^{2}$,
while on the right-hand side we have
\be
iq_{\mu}\langle\pi^{-}\eta^{(\prime)}|\bar{d}\gamma^{\mu}u|0\rangle=
ic^{V}_{\pi^{-}\eta^{(\prime)}}\left[(m_{\eta^{(\prime)}}^{2}-m_{\pi^{-}}^{2})
F_{+}^{\pi^{-}\eta^{(\prime)}}(s)-sF_{-}^{\pi^{-}\eta^{(\prime)}}(s)\right]\ .
\label{divergence2}
\ee
Then, by equating Eqs.~(\ref{divergence1}) and (\ref{divergence2}), we link
$F_{-}^{\pi^{-}\eta^{(\prime)}}(s)$ with $F_{0}^{\pi^{-}\eta^{(\prime)}}(s)$ through
\be
F_{-}^{\pi\eta^{(\prime)}}(s)=-\frac{\Delta_{\pi^{-}\eta^{(\prime)}}}{s}\left[\frac{c^{S}_{\pi\eta^{(\prime)}}}
{c^{V}_{\pi\eta^{(\prime)}}}\frac{\Delta^{\rm QCD}_{K^{0}K^{+}}}{\Delta_{\pi^{-}\eta^{(\prime)}}}F_{0}^{\pi\eta^{(\prime)}}(s)+
F_{+}^{\pi^{-}\eta^{(\prime)}}(s)\right]\ ,
\label{F-}
\ee
and the hadronic matrix element finally reads 
\be
\begin{array}{rl}
\langle \pi^{-}\eta^{(\prime)}|\bar{d}\gamma^{\mu}u|0\rangle=&\!\!\!
\displaystyle{c^{V}_{\pi\eta^{(\prime)}}\left[(p_{\eta^{(\prime)}}-p_{\pi})^{\mu}+
\frac{\Delta_{\pi^{-}\eta^{(\prime)}}}{s}q^{\mu}\right]
F_{+}^{\pi\eta^{(\prime)}}(s)}\\[2ex]
&\!\!\!
\displaystyle{+\,c^{S}_{\pi^{-}\eta^{(\prime)}}
\frac{\Delta^{\rm QCD}_{K^{0}K^{+}}}{s}q^{\mu}F_{0}^{\pi^{-}\eta^{(\prime)}}(s)}\ .
\end{array}
\label{vectorcurrent}
\ee
The advantage of the parameterization as given in Eq.~(\ref{vectorcurrent}) is that the vector(scalar) form factor
$F_{+(0)}^{\pi^{-}\eta^{(\prime)}}(s)$ is in direct correspondence with the final $P(S)$-wave state, respectively.
Moreover, the finiteness of the matrix element at the origin imposes\footnote{We
will come back to Eq.~(\ref{VFF0}) in Sect.~\ref{predictions} in order to check the consistency of our input values.}
\be
F_+^{\pi^-\eta^{(\prime)}}(0)=
-\frac{c^S_{\pi^-\eta^{(\prime)}}}{c^V_{\pi^-\eta^{(\prime)}}}
\frac{\Delta^{\rm QCD}_{K^0K^+}}{\Delta_{\pi^-\eta^{(\prime)}}}F_0^{\pi^-\eta^{(\prime)}}(0)\ .
\label{VFF0} 
\ee
Therefore, the differential decay width of the $\tau^{-}\to\pi^{-}\eta^{(\prime)}\nu_{\tau}$ decay
as a function of the invariant mass of the $\pi^{-}\eta^{(\prime)}$ system can be written as
\be
\begin{array}{l}
\displaystyle{\frac{d\Gamma\left(\tau^-\to\pi^-\eta^{(\prime)}\nu_\tau\right)}{d\sqrt{s}}=
\frac{G_F^2M_\tau^3}{24\pi^3s}S_{\rm EW}|V_{ud}F_+^{\pi^-\eta^{(\prime)}}(0)|^2
\left(1-\frac{s}{M_\tau^2}\right)^2}\\[2ex]
\qquad
\displaystyle{\times\left[\left(1+
\frac{2s}{M_\tau^2}\right)q_{\pi^-\eta^{(\prime)}}^3(s)|\widetilde{F}_+^{\pi^-\eta^{(\prime)}}(s)|^2+
\frac{3\Delta_{\pi^-\eta^{(\prime)}}^2}{4s}q_{\pi^-\eta^{(\prime)}}(s)|\widetilde{F}_0^{\pi^-\eta^{(\prime)}}(s)|^2\right]}\ ,
\end{array}
\label{width}
\ee
where $q_{PQ}(s)=\sqrt{s^2-2s\Sigma_{PQ}+\Delta_{PQ}^2}/2\sqrt{s}$, $\Sigma_{PQ}=m_P^2+m_Q^2$ and
\be
\widetilde{F}_{+,0}^{\pi^-\eta^{(\prime)}}(s)=\frac{F_{+,0}^{\pi^-\eta^{(\prime)}}(s)}{F_{+,0}^{\pi^-\eta^{(\prime)}}(0)}\ ,
\ee
are the two form factors normalized to unity at the origin.
They encode the unknown strong dynamics occurring in the transition.
Their descriptions will be given in Secs.~\ref{VectorFormFactor} and \ref{ScalarFormFactors}, respectively.
Regarding the global pre-factors,
we employ $S_{\rm EW}=1.0201$~\cite{Erler:2002mv}, accounting for short-distance electroweak corrections,
and $V_{ud}=0.97425(8)(10)(18)$ \cite{pdg},
while the normalization $F_{+}^{\pi^{-}\eta^{(\prime)}}(0)$ is an isospin-violating quantity of $\mathcal{O}(m_{d}-m_{u})$,
whose value will be deduced in the next section,
which brings an overall suppression explaining the smallness of the corresponding decay widths.
In fact, in the limit of exact isospin, $m_{u}=m_{d}$ and $e=0$, $F_{+}^{\pi^{-}\eta^{(\prime)}}(0)=0$
and these processes would be forbidden in the SM.

\section{$\pi^{-}\eta^{(\prime)}$ Vector Form Factor}
\label{VectorFormFactor}
We derive the $\pi^{-}\eta^{(\prime)}$ vector form factor within the context of resonance chiral theory 
(RChT)~\cite{Ecker:1988te},
which extends chiral perturbation theory \cite{ChPT} by adding resonances as explicit degrees of freedom.
A short introduction to the topic can be found in Ref.~\cite{Portoles:2010yt},
where references concerning its varied phenomenological applications are given.
In Refs.~\cite{Escribano:2013bca,Escribano:2014pya}
we have also provided a short review of the theory as applied to the computation of the vector and scalar
$K^{-}\eta^{(\prime)}$ form factors describing the decays $\tau^{-}\to K^{-}\eta^{(\prime)}\nu_{\tau}$.
In the present analysis, we would occasionally refer the interested reader to the former references
though some comments will be given in the following for consistency.

It is not straightforward to incorporate the dynamics of the $\eta$ and $\eta^{\prime}$ mesons in a chiral framework
(see, for instance, Ref.~\cite{Schechter:1992iz}).
The pseudoscalar singlet $\eta_0$ is absent in $SU(3)$ ChPT and their effects are encoded in the next-to-leading order
low-energy constant $L_7$.
To take into account consistently the effects of the singlet in an explicit way one must perform a simultaneous expansion
not only in terms of momenta $(p^2)$ and quark masses $(m_q)$ but also in the number of colors $(1/N_c)$.
In this framework, known as Large-$N_c$ ChPT~\cite{Kaiser:2000gs},
the singlet becomes a ninth pseudo-Goldstone boson and the $\eta$-$\eta^\prime$ mixing
can be understood in a perturbative manner\footnote{In
this simultaneous expansion the chiral loops are counted as next-to-next-to-leading order corrections and
thus considered negligible \cite{Kaiser:2000gs}.
This fact is in part corroborated numerically.}.
At lowest order, the physical states $(\eta,\eta^\prime)$ are related to the mathematical states $(\eta_8,\eta_0)$
in the so-called octet-singlet basis by a simple two-dimensional rotation matrix involving one single mixing angle.
At the same order, the four different decay constants related to the $\eta$-$\eta^\prime$ system are all equal
to the pion decay constant in the chiral limit.
At next-to-leading order, however, besides mass-matrix diagonalization one requires to perform first
a wave-function renormalisation of the fields due to the non-diagonal form of the kinetic term of the Lagrangian.
This two-step procedure makes the single mixing angle at lowest order to be split in two mixing angles
at next-to-leading order\footnote{For
a detailed explanation of the two-mixing angle scheme in the large-$N_c$ ChPT at next-to-leading order
in the octet-singlet basis, see, for instance, the appendix B in Ref.~\cite{Escribano:2010wt}.
Several phenomenological analyses using this basis or the so-called quark-flavour basis are
Refs.~\cite{Feldmann:1998vh,Feldmann:1998sh,Escribano:2005qq}.
Other comprehensive reviews are Refs.~\cite{Feldmann:2002kz,Feldmann:1999uf}.}.
The magnitude of this splitting is given in the octet-singlet basis by the difference of the $F_K$ and $F_\pi$
decay constants, that is, a $SU(3)$-breaking correction \cite{Kaiser:1998ds}.
At this order, now, the decay constants are all different due to these wave-function--renormalisation corrections.
Being this two-mixing angle scheme unavoidable at next-to-leading order in the large-$N_c$ chiral expansion,
one can express their associated parameters either in the form of
two mixing angles $(\theta_8, \theta_0)$ and two decay constants $(f_8,f_0)$
or one mixing angle, the one appearing at lowest order, and three wave-function--renormalisation corrections,
appearing only at next-to-leading order.
In this work, we will follow the second option.
Needless to say, the mixing so far involves only the $\eta$ and $\eta^\prime$ mesons in the isospin limit,
but if isospin symmetry is broken, as it is our case, the $\pi^0$ is also involved,
and instead of using one mixing angle and three wave-function--renormalisation corrections
we will need to use three lowest order mixing angles,
$\theta_{\eta\eta^\prime}$ for the $\eta$-$\eta^\prime$, $\theta_{\pi\eta}$ for the $\pi$-$\eta$ and
$\theta_{\pi\eta^\prime}$ for the $\pi$-$\eta^\prime$ systems, respectively, and
the corresponding six wave-function--renormalisation corrections.
Since we are in the context of RChT, these wave-function--renormalisation corrections are assumed to be saturated
by the exchange of a nonet of scalar resonances and therefore expressed in terms of the associated $c_d$ and $c_m$
coupling constants (see below).

Because the size of isospin-breaking corrections due to the light-quark mass difference are given in terms of the ratio
$(m_d-m_u)/m_s$ and hence very small, the two former mixing angles involving the $\pi^0$ can be well approximated by
their Taylor expansion at first order.
Then, the orthogonal matrix connecting the mathematical and physical states at lowest order can be written as
\be
\left(
\begin{array}{c}
\pi^0\\
\eta\\
\eta^\prime
\end{array}
\right)=
\left(
\begin{array}{ccc}
1&
\varepsilon_{\pi\eta}{\rm c}\theta_{\eta\eta^\prime}+\varepsilon_{\pi\eta^\prime}{\rm s}\theta_{\eta\eta^\prime}&
\varepsilon_{\pi\eta^\prime}{\rm c}\theta_{\eta\eta^\prime}-\varepsilon_{\pi\eta}{\rm s}\theta_{\eta\eta^\prime}\\
-\varepsilon_{\pi\eta}&{\rm c}\theta_{\eta\eta^\prime}&-{\rm s}\theta_{\eta\eta^\prime}\\
-\varepsilon_{\pi\eta^\prime}&{\rm s}\theta_{\eta\eta^\prime}&{\rm c}\theta_{\eta\eta^\prime}
\end{array}
\right)\cdot
\left(
\begin{array}{c}
\pi_3\\
\eta_8\\
\eta_0
\end{array}
\right)\ ,
\label{mixing}
\ee
where $\varepsilon_{\pi\eta^{(\prime)}}$ are the approximated $\pi^0$-$\eta^{(\prime)}$ mixing angles
and $({\rm c},{\rm s})\equiv (\cos,\sin)$.
Using this parametrization for the rotation matrix,
we preserve the common $\eta$-$\eta^\prime$ mixing description,
when both $\varepsilon_{\pi\eta^{(\prime)}}$ are fixed to 0,
and the one for $\pi$-$\eta^{(\prime)}$ mixing,
when both $\theta_{\eta\eta^\prime}$ and $\varepsilon_{\pi\eta^{\prime()}}$ are set to 0.
A detailed illustration of this $\pi^0$-$\eta$-$\eta^\prime$ mixing can be found in Ref.~\cite{Kroll:2005sd},
from where we borrow the numerical values
$\hat\varepsilon_{\pi\eta}\equiv\varepsilon_{\pi\eta}(z=0)=0.017(2)$ and
$\hat\varepsilon_{\pi\eta^\prime}\equiv\varepsilon_{\pi\eta^\prime}(z=0)=0.004(1)$ as a check of our results.
For the $\eta$-$\eta^\prime$ mixing angle we take
$\theta_{\eta\eta^\prime}=(-13.3\pm 0.5)^\circ$ \cite{Ambrosino:2009sc}\footnote{In
Ref.~\cite{Ambrosino:2009sc},
the value $\phi_{\eta\eta^\prime}=(41.4\pm 0.5)^\circ$ is obtained in the quark-flavor basis.
However, at lowest order, this value is equivalent in the octet-singlet basis to
$\theta_{\eta\eta^\prime}=\phi_{\eta\eta^\prime}-\arctan\sqrt{2}=(-13.3\pm 0.5)^\circ$.}.

As stated before, the $\pi^{-}\eta^{(\prime)}$ VFFs will be calculated in the framework of RChT.
There are four different types of contributions in total.
At leading order, there is the contribution from the lowest order of large-$N_c$ ChPT.
At next-to-leading order, there are, in addition, 
the contribution from the exchange of explicit vector resonances,
the so-called vacuum insertions
and the wave-function--renormalisation contributions.
The latter two are written in terms of the explicit exchange of scalar resonances
and seen to cancel each other \cite{Oller:2000ug}.
As a result, we obtain
\be
F_{+}^{\pi^{-}\eta^{(\prime)}}(s)=\varepsilon_{\pi\eta^{(\prime)}}
\left(1+\sum_{V}\frac{F_{V}G_{V}}{F^{2}}\frac{s}{M_{V}^{2}-s}\right)\ ,
\label{VFFpietaetap}
\ee
where the prefactor denotes it occurs via $\pi^0$-$\eta$-$\eta^\prime$ mixing
and the parenthesis includes the direct contact term plus the exchange of an infinite number
of vector resonances organized in nonets\footnote{At
leading order in $1/N_c$ at this stage, \textit{i.e.}, with an infinite number of zero-width resonances \cite{'tHooft:1973jz}.}
($F_V$ and $G_V$ are the two coupling constants of the Lagrangian of one nonet of vectors
coupled to pseudoscalars, $M_V$ the common nonet vector mass, and $F$ the pion decay constant in the chiral limit).
%Once the QCD asymptotic behavior of the form factors is imposed,
%that is, they are ${\cal O}(1/s)$ for large $s$, which implies $\sum_V F_V G_V=F^2$,
%these can be finally written as
%\be
%F_{+}^{\pi^{-}\eta^{(\prime)}}(s)=\varepsilon_{\pi\eta^{(\prime)}}\frac{M_{V}^{2}}{M_{V}^{2}-s}\ .
%\label{VFFpietaetapQCD}
%\ee

\begin{figure}
\centering
\includegraphics[scale=0.75]{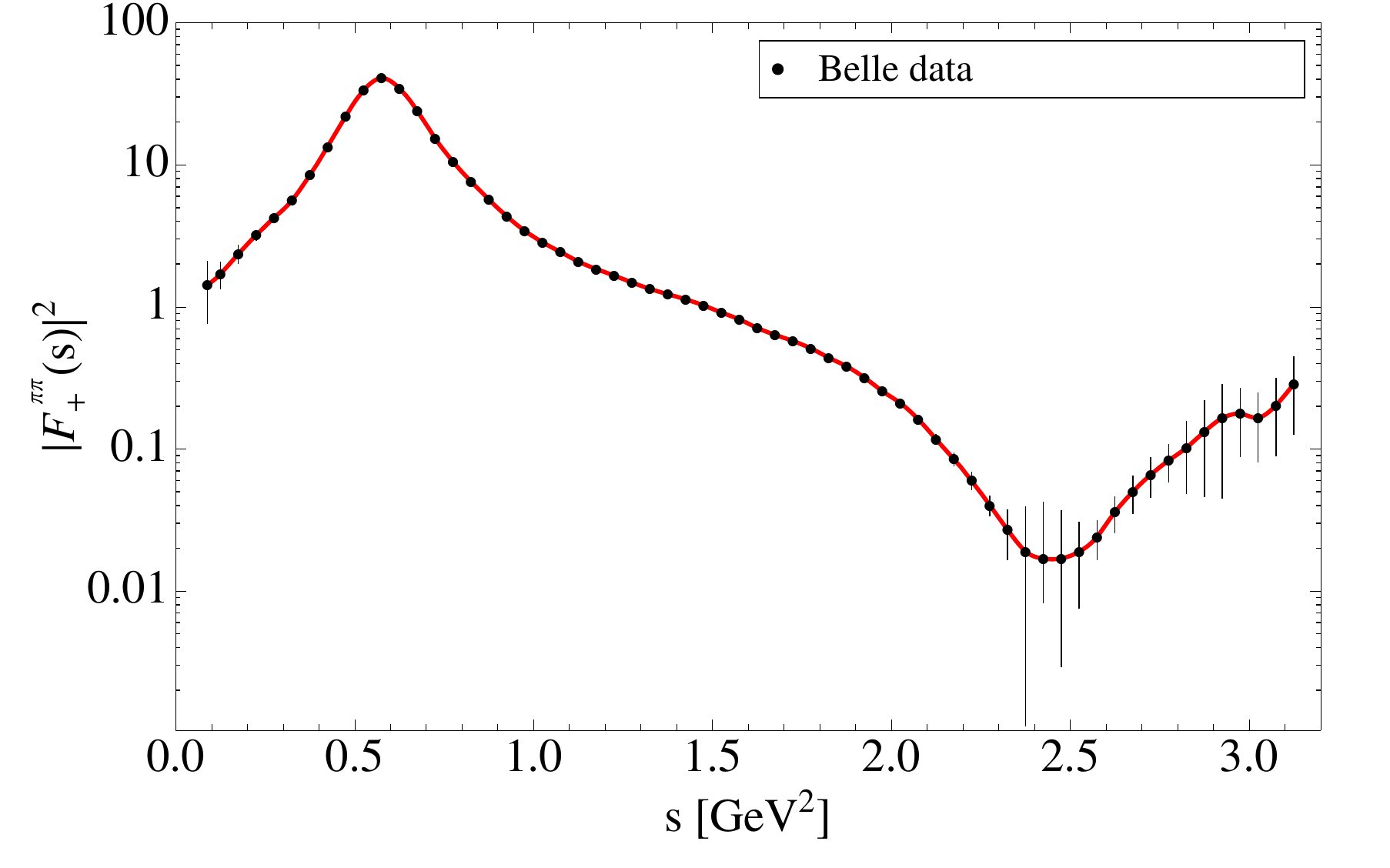}
\caption{
$\pi^{-}\pi^{0}$ vector form factor as obtained by the Belle Collaboration \cite{Fujikawa:2008ma} (black circles).
The red solid curve is an interpolation of these data.}
\label{VFFpipi}
\end{figure}

Interestingly, the term in parenthesis appearing in Eq.~(\ref{VFFpietaetap})
is nothing but what one would have obtained if the $\pi^{-}\pi^{0}$ VFF had been computed instead.
Hence, written in this way, the $\pi^{-}\eta^{(\prime)}$ VFFs are given in terms of the well-known $\pi^{-}\pi^{0}$ VFF
(see, for instance, Refs.~\cite{Dumm:2013zh,Guerrero:1997ku} for a review).
Their value at the origin are $F_{+}^{\pi^{-}\eta^{(\prime)}}(0)=\varepsilon_{\pi\eta^{(\prime)}}$,
and as a consequence the normalized form factors are both the same and equal to the normalized $\pi^{-}\pi^{0}$ one,
that is
\be
\widetilde{F}_{+}^{\pi^{-}\eta}(s)=\widetilde{F}_{+}^{\pi^{-}\eta^{\prime}}(s)=\widetilde{F}_{+}^{\pi^{-}\pi^{0}}(s)\ .
\label{VFFnormalised}
\ee
The above relation allows us to implement the well-known experimental data on the $\pi^{-}\pi^{0}$ VFF
to describe the $\pi^{-}\eta^{(\prime)}$ decay modes we are interested in.
In particular, we employ the latest experimental determination obtained by the Belle Collaboration
from the measurement of the decay $\tau^{-}\to\pi^{-}\pi^{0}\nu_{\tau}$\footnote{The
contribution of the scalar form factor entering into the $\pi^{-}\pi^{0}$ decay mode is weighted by
$\Delta_{\pi^-\pi^0}^2$, thus heavily suppressed by isospin \cite{Cirigliano:2001er} and usually neglected.},
which is shown in Fig.~\ref{VFFpipi} (the set of data is borrowed from the Table VI of Ref.~\cite{Fujikawa:2008ma}).
In this manner, we are not only taking into account the dominant vector resonant contribution given by the $\rho(770)$,
whose effect is clearly seen from the neat peak around $0.6$ GeV$^{2}$,
but also the effects of higher radial excitations such as the $\rho^{\prime}(1450)$ and $\rho^{\prime\prime}(1700)$
(see their manifestation in the form of a negative interference with the $\rho$ in the energy region between
2 and 3 GeV$^{2}$).
An interesting check would be then to compare these data with theoretical descriptions of this form factor,
such as the ones given by dispersion relations, where the contributions of the different states can be switched on and off,
to discern the number of participating resonances \cite{Dumm:2013zh, Celis:2013xja}.

\section{$\pi^{-}\eta^{(\prime)}$ Scalar Form Factor}
\label{ScalarFormFactors}
Any description of a physical observable involving light scalar mesons has been always controversial\footnote{See
{\textit{e.g.}} the ``Note on scalar mesons below 2 GeV'' in Ref.~\cite{pdg} for a review.},
and simple model parameterizations do not typically succeed.
In this work, in order to construct a reasonable description of the participant scalar form factors
we will basically exploit two powerful theoretical arguments:
the required analytical structure of the form factor and the unitarity of the scattering matrix.
In what follows, we will tackle three different parameterizations in increasing degree of completeness.

\subsection{Breit-Wigner approach}
\label{BWdescription}
Our initial approach for describing the required $\pi^{-}\eta^{(\prime)}$ scalar form factor (SFF)
is, as in the case of the VFF, the RChT framework.
In the large-$N_c$ limit, the octet of scalar resonances and the singlet become degenerate in the chiral limit
(with common mass $M_S$), and all them are collected in a nonet.
The calculation of these SFFs is performed again at next-to-leading order in the simultaneous expansion
in terms of momenta and the number of colors, and the different contributions to them are
the lowest order one from large-$N_c$ ChPT and the three next-to-leading order ones from RChT,
which are, in order, the explicit exchange of scalar resonances, the vacuum insertions,
and the wave-function--renormalisation contributions.
The resulting SFFs are\footnote{As
a starting point, we assume there is only a nonet of scalar resonances.
Later on, we will include a second one.
Moreover, we use in the calculation of the form factors isospin-averaged $\pi(K)$ masses $m_{\pi(K)}$
which will be in the following identified as their corresponding charged masses,
being the differences higher-order isospin-breaking corrections.}
\be
\begin{array}{l}
F_{0}^{\pi^{-}\eta^{(\prime)}}(s)=c_{0}^{\pi^{-}\eta^{(\prime)}}
\displaystyle{\left[1-\frac{8c_{m}(c_{m}-c_{d})}{{F}^{2}}\frac{2m_{K}^{2}-m_{\pi}^{2}}{M_{S}^{2}}\right.}\\[4ex]
\qquad
\displaystyle{\left.+\frac{4c_{m}}{F^{2}}
\frac{(c_{m}-c_{d})2m_{\pi}^{2}+c_{d}\left(s+m_{\pi}^{2}-m_{\eta^{(\prime)}}^{2}\right)}{M_{S}^{2}-s}\right]}\ ,
\end{array}
\label{formfactorRChT}
\ee
where $c_{0}^{\pi^{-}\eta}=\cos\theta_{\eta\eta^\prime}-\sqrt{2}\sin\theta_{\eta\eta^\prime}$
and $c_{0}^{\pi^{-}\eta^\prime}=\cos\theta_{\eta\eta^\prime}+\sin\theta_{\eta\eta^\prime}/\sqrt{2}$
for the $\pi\eta$ and $\pi\eta^\prime$ channels, respectively,
and $c_{d(m)}$ are the couplings appearing in the derivative(mass) terms of the Lagrangian
involving the nonets of scalar and pseudoscalar mesons.
A similar analysis was done in Ref.~\cite{Jamin:2001zq} for the $K\pi$, $K\eta$ and $K\eta^\prime$ SFFs.
Once the QCD asymptotic behavior of the form factors is imposed,
that is, they are ${\cal O}(1/s)$ for large $s$, which implies $c_d-c_m=0$ and $4c_d c_m=F^2$,
and hence $c_d=c_m=F/2$ \cite{Jamin:2001zq},
these can be finally written as \cite{Roig:2013ila}
\be
F_{0}^{\pi^{-}\eta^{(\prime)}}(s)=c_{0}^{\pi^{-}\eta^{(\prime)}}
\left(1+\frac{\Delta_{\pi^{-}\eta^{(\prime)}}}{M_{S}^{2}}\right)\frac{M_{S}^{2}}{M_{S}^{2}-s}\ ,
\label{B_W}
\ee
and their value at the origin are
\be
F_{0}^{\pi^{-}\eta^{(\prime)}}(0)=c_{0}^{\pi^{-}\eta^{(\prime)}}
\left(1+\frac{\Delta_{\pi^{-}\eta^{(\prime)}}}{M_{S}^{2}}\right)\ .
\label{SFF0}
\ee
These normalizations can now be incorporated into Eq.~(\ref{VFF0})
to give a prediction of the normalizations of the related VFFs:
\be
\begin{array}{rcl}
F_{+}^{\pi^{-}\eta}(0)&=&
\displaystyle{-\frac{\cos\theta_{\eta\eta^\prime}-\sqrt{2}\sin\theta_{\eta\eta^\prime}}{\sqrt{3}}
\frac{\Delta^{\rm QCD}_{K^{0}K^{+}}}{\Delta_{\pi^{-}\eta}}
\left(1+\frac{\Delta_{\pi^{-}\eta}}{M_{S}^{2}}\right)}\\[4ex]
&=&\cos\phi_{\eta\eta^\prime}
\displaystyle{\frac{m_{K^0}^2-m_{K^+}^2-m_{\pi^0}^2+m_{\pi^+}^2}{m_{\eta}^2-m_{\pi^{-}}^2}
\left(1-\frac{m_{\eta}^2-m_{\pi^{-}}^2}{M_{S}^{2}}\right)}\ ,
\end{array}
\label{SFF0eta}
\ee
and
\be
\begin{array}{rcl}
F_{+}^{\pi^{-}\eta^\prime}(0)&=&
\displaystyle{-\frac{\sin\theta_{\eta\eta^\prime}+\sqrt{2}\cos\theta_{\eta\eta^\prime}}{\sqrt{3}}
\frac{\Delta^{\rm QCD}_{K^{0}K^{+}}}{\Delta_{\pi^{-}\eta^{\prime}}}
\left(1+\frac{\Delta_{\pi^{-}\eta^{\prime}}}{M_{S}^{2}}\right)}\\[4ex]
&=&\sin\phi_{\eta\eta^\prime}
\displaystyle{\frac{m_{K^0}^2-m_{K^+}^2-m_{\pi^0}^2+m_{\pi^+}^2}{m_{\eta^\prime}^2-m_{\pi^{-}}^2}
\left(1-\frac{m_{\eta^\prime}^2-m_{\pi^{-}}^2}{M_{S}^{2}}\right)}\ ,
\end{array}
\label{SFF0etaprime}
\ee
where the $\eta$-$\eta^\prime$ mixing has been expressed for simplicity in the quark-flavor basis,
$\cos\phi_{\eta\eta^\prime}=(\cos\theta_{\eta\eta^\prime}-\sqrt{2}\sin\theta_{\eta\eta^\prime})/\sqrt{3}$ and
$\sin\phi_{\eta\eta^\prime}=(\sin\theta_{\eta\eta^\prime}+\sqrt{2}\cos\theta_{\eta\eta^\prime})/\sqrt{3}$, and
$\Delta^{\rm QCD}_{K^{0}K^{+}}=m_{K^0}^2-m_{K^+}^2-{\Delta m_K^2}_{\rm elm}=
m_{K^0}^2-m_{K^+}^2-m_{\pi^0}^2+m_{\pi^+}^2$
has been estimated from the $K^0$-$K^+$ mass difference corrected for mass contributions of electromagnetic origin
according to Dashen's theorem \cite{Dashen:1969eg,Langacker:1974nm}.
Comparing these VFFs normalizations with those obtained after Eq.~(\ref{VFFpietaetap}),
one finally gets
\be
\varepsilon_{\pi\eta^{(\prime)}}=
\cos\phi_{\eta\eta^\prime}(\sin\phi_{\eta\eta^\prime})
\frac{m_{K^0}^2-m_{K^+}^2-m_{\pi^0}^2+m_{\pi^+}^2}{m_{\eta^{(\prime)}}^2-m_{\pi^{-}}^2}
\left(1-\frac{m_{\eta^{(\prime)}}^2-m_{\pi^{-}}^2}{M_{S}^{2}}\right)\ ,
\label{epsilonetaetaprime}
\ee
for the $\pi\eta$ and $\pi\eta^\prime$ cases, respectively.
It is worth noticing that the former equation is equivalent up to higher-order isospin corrections
to Eq.~(31) in Ref.~\cite{Kroll:2005sd} after the identification
$z\equiv(f_u-f_d)/(f_u+f_d)=-(m_{K^0}^2-m_{K^+}^2-m_{\pi^0}^2+m_{\pi^+}^2)/M_{S}^{2}$.
The former equality allows for an estimate of this parameter, $z\simeq -5\times 10^{-3}$ for $M_S=980$ MeV,
in agreement with the conclusion in Ref.~\cite{Kroll:2005sd} that $z<0.015$.
From Eq.~(\ref{epsilonetaetaprime}), we can also provide a numerical determination of the
$\pi\eta^{(\prime)}$ mixing angles,
$\varepsilon_{\pi\eta}=(9.8\pm 0.3)\times 10^{-3}$
and
$\varepsilon_{\pi\eta^{\prime}}=(2.5\pm 1.5)\times 10^{-4}$,
which are far, specially in the latter case, from their infinite scalar mass limit,
$\hat\varepsilon_{\pi\eta}\equiv\varepsilon_{\pi\eta}(M_S\to\infty)=0.014$
and
$\hat\varepsilon_{\pi\eta^{\prime}}\equiv\varepsilon_{\pi\eta^{\prime}}(M_S\to\infty)=0.0038$,
in accordance with Ref.~\cite{Feldmann:1998sh}.
These values were calculated using $\phi_{\eta\eta^\prime}=(41.4\pm 0.5)^\circ$ \cite{Ambrosino:2009sc}.
As seen, $\varepsilon_{\pi\eta^{\prime}}$ is one order of magnitude smaller than $\hat\varepsilon_{\pi\eta^{\prime}}$
caused by the strong suppression due to $m_{\eta^\prime}\simeq M_S$.

\begin{figure}
\centering
\includegraphics[scale=0.75]{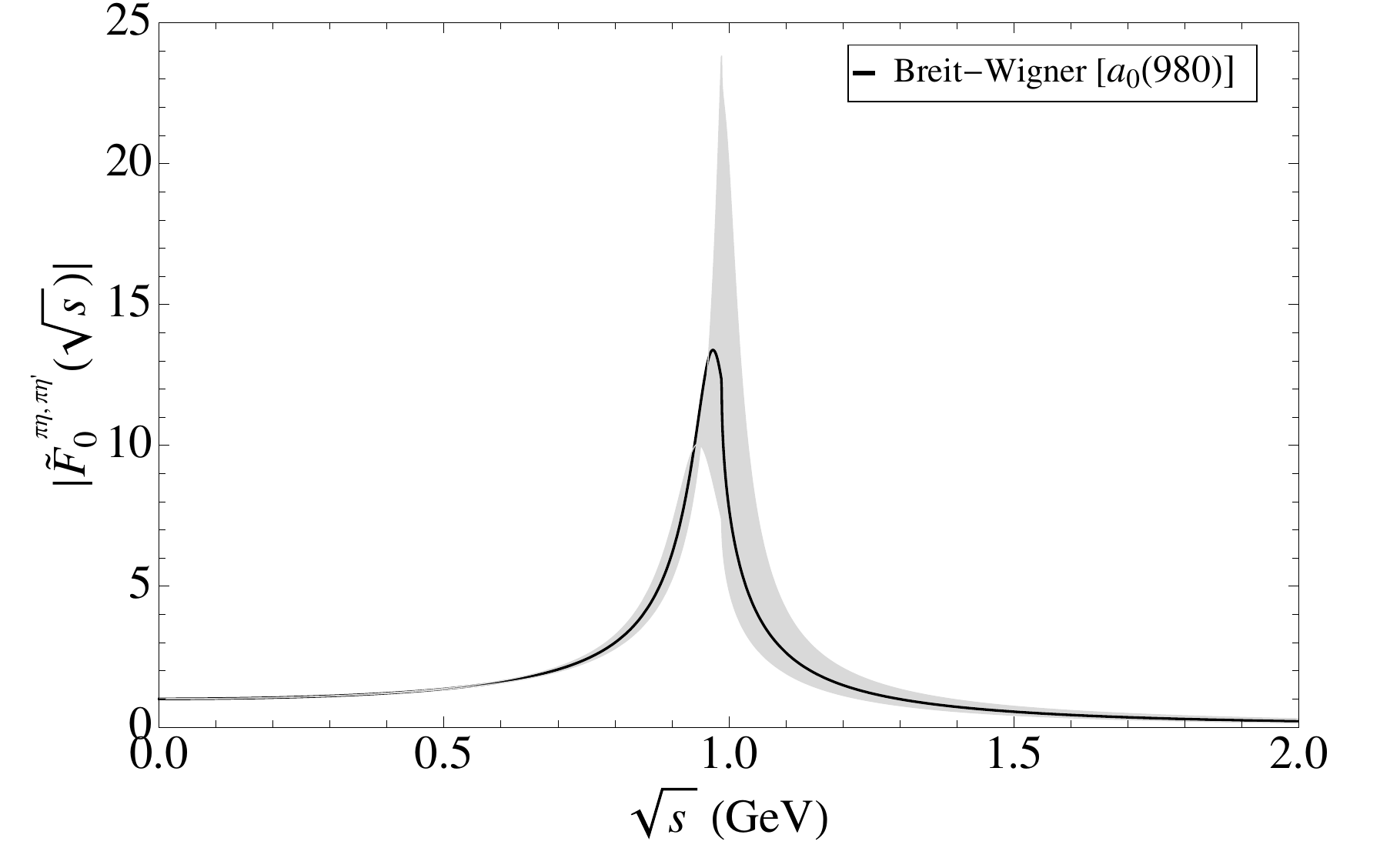}
\caption{Normalized
$\pi^{-}\eta^{(\prime)}$ scalar form factors as obtained from the Breit-Wigner approach
described in Sec.~\ref{BWdescription}.
The gray error band accounts for the (uncorrelated) uncertainty on the mass and width
of the $a_{0}(980)$ resonance.}
\label{BWplot}
\end{figure}

The description of the SFFs in the form of Eq.~(\ref{B_W}) begins to fail in the vicinity of the resonance region.
It breaks down for $s=M_{S}^2$ which corresponds to an on-shell intermediate scalar resonance.
A common and simple way to cure this limitation is by promoting the scalar propagator
$1/(M_{S}^2-s)$ to $1/(M_{S}^2-s-iM_{S}\Gamma_{S}(s))$,
where the corresponding energy-dependent width computed within RChT in this case reads
\be
\Gamma_{S}(s)=\Gamma_{S}(M_{S}^{2})\left(\frac{s}{M_{S}^{2}}\right)^{3/2}\frac{h(s)}{h(M_{S}^{2})}\ ,
\ee
with ($\sigma_{PQ}(s)=2q_{PQ}(s)/\sqrt{s}\times\Theta(s-(m_P+m_Q)^2)$ is a kinematical factor)
\be
\begin{array}{l}
h(s)=
\displaystyle{\sigma_{K^-K^0}(s)
+2\cos^2\phi_{\eta\eta^\prime}\left(1+\frac{\Delta_{\pi^{-}\eta}}{s}\right)^{2}\sigma_{\pi^{-}\eta}(s)}\\[4ex]
\qquad\quad
\displaystyle{
+2\sin^2\phi_{\eta\eta^\prime}\left(1+\frac{\Delta_{\pi^{-}\eta^{\prime}}}{s}\right)^{2}\sigma_{\pi^{-}\eta^{\prime}}(s)}\ ,
\end{array}
\ee
for the $a_0(980)$ resonance case coupling dominantly to the $\pi\eta$ system\footnote{Current
understanding favors that the meson multiplet including this resonance does not survive in the large-$N_c$ limit
(see e.g.~Refs.~\cite{Pelaez:2003dy,Cirigliano:2003yq,Caprini:2005zr,DescotesGenon:2006uk}).
However, since this Breit-Wigner--like model is only considered for illustrative purposes this fact will be ignored
as it is usually done in this approach.}.
In this way, we have incorporated into our description some elastic and inelastic unitarity corrections through
resumming the imaginary part of the $\pi^{-}\eta^{(\prime)}$ and $K^{-}K^{0}$ self-energy loop insertions
into the propagator, accounting for rescattering effects of the final state hadrons.
Nonetheless, this description is not strictly unitary neither in its elastic form
(since we have accommodated inelasticities into the description) nor in an inelastic fashion
which would require to couple the channels in a more elaborated way.
In addition, this description is neither fully analytic in the sense that the real part of the loop functions has been neglected.
Usually, this option, known as the Breit-Wigner (BW) representation, is widely used in the literature even though
it might not be an appropriate choice for describing data
(as we have pointed out in Refs.~\cite{Escribano:2013bca,Roig:2014mva}).
Notwithstanding, we have considered interesting to discuss it as a starting point.
Using the values $M_{S}=(980\pm 20)$ MeV and $\Gamma_{S}=(75\pm 25)$ MeV \cite{pdg}
for the BW-mass and -width of the $a_0(980)$ resonance,
the SFFs at the origin, see Eq.~(\ref{SFF0}), are predicted to be
$F_0^{\pi\eta}(0)=0.92\pm 0.02$ and $F_0^{\pi\eta^{\prime}}(0)=0.05\pm 0.03$, respectively.resuming
Once these normalizations are taken into account, the resulting normalized SFFs are identical
in the RChT framework, that is, $\widetilde F_{0}^{\pi^{-}\eta}(s)=\widetilde F_{0}^{\pi^{-}\eta^\prime}(s)$.
In Fig.~\ref{BWplot}, we provide their graphical account by considering $a_{0}(980)$
as the mediated scalar resonance. 

\begin{figure}
\centering
\includegraphics[scale=0.4]{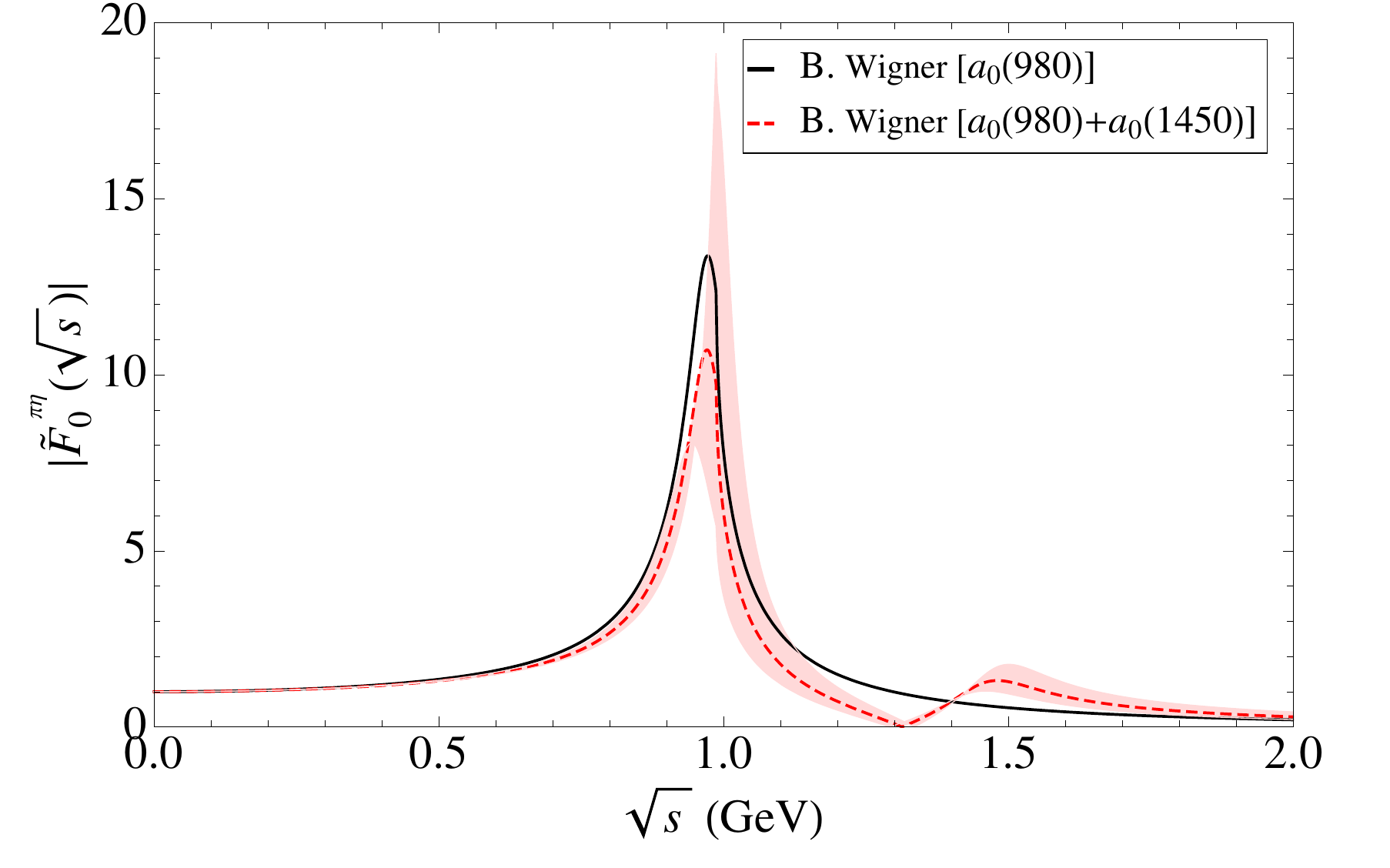}
\includegraphics[scale=0.4]{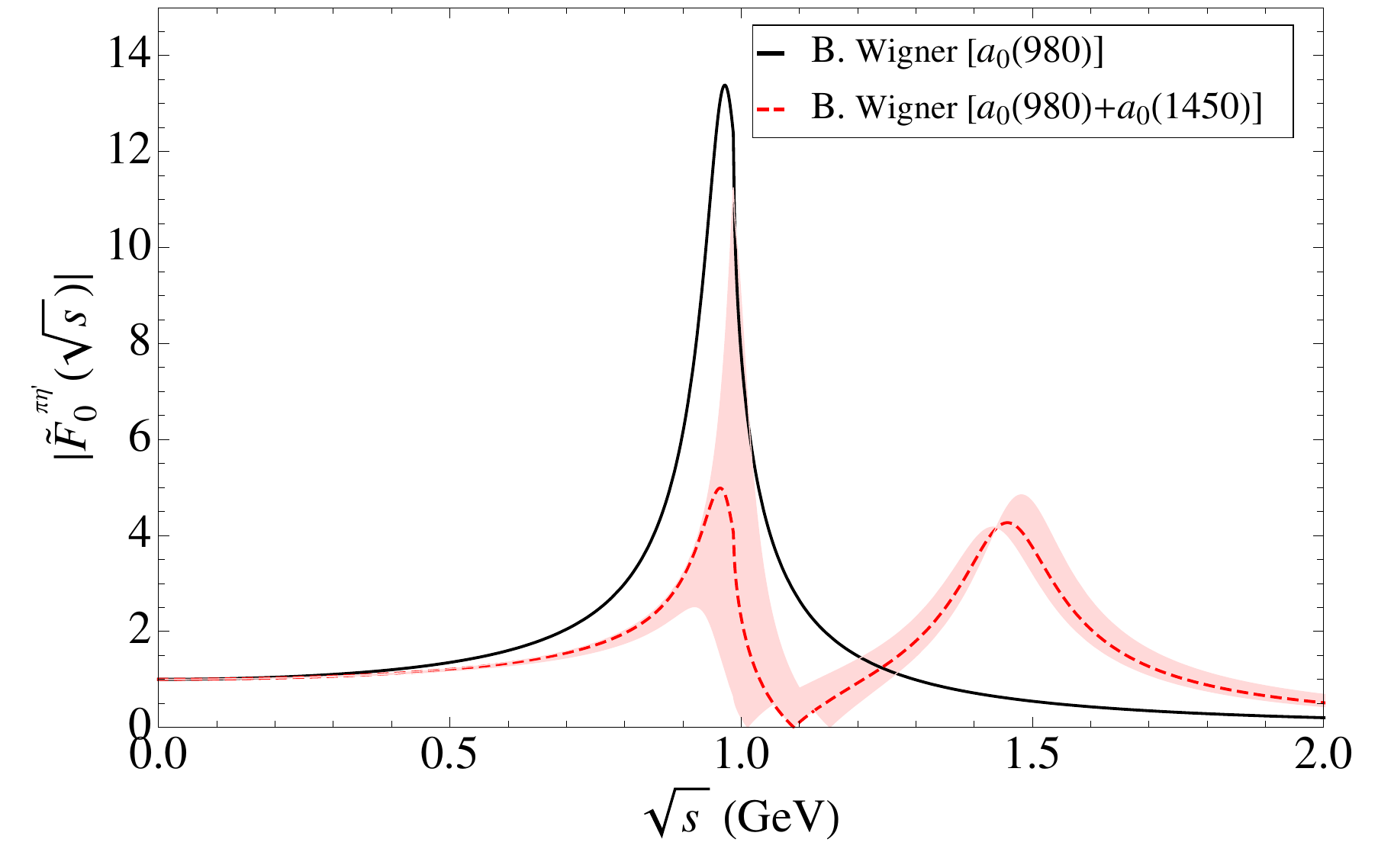}
\caption{Normalized
$\pi^{-}\eta$ (left plot) and $\pi^{-}\eta^{\prime}$ (right plot) scalar form factors
as obtained from the Breit-Wigner approach described in Sec.~\ref{BWdescription}
including two resonances (red dashed curves) or a single resonance (solid black curves).
The red error bands account for the (uncorrelated) uncertainty on the mass and width of the
$a_{0}(980)$ and $a_{0}(1450)$ resonances.}
\label{BWplot2}
\end{figure}

The above description can be generalized to take into consideration further resonances
with the same quantum numbers of the $a_{0}(980)$.
In particular, we will also include the $a_{0}(1450)$ resonance whose effects, in spite of its mass, could be noticeable
within the available phase space.
For the same reason, however, no more resonances will be considered henceforth.
The SFFs in the framework of RChT including two resonances then read as
\be
\begin{array}{l}
F_{0}^{\pi^{-}\eta^{(\prime)}}(s)=c_{0}^{\pi^{-}\eta^{(\prime)}}\\[2ex]
\displaystyle{\times\left[1-\frac{8c_{m}(c_{m}-c_{d})}{{F}^{2}}\frac{2m_{K}^{2}-m_{\pi}^{2}}{M_{S}^{2}}
+\frac{4c_{m}}{F^{2}}\frac{(c_{m}-c_{d})2m_{\pi}^{2}+c_{d}
\left(s+m_{\pi}^{2}-m_{\eta^{(\prime)}}^{2}\right)}{M_{S}^{2}-s}\right.}\\[4ex]
\quad
\displaystyle{\left.-\frac{8c^{\prime}_{m}(c^{\prime}_{m}-c^{\prime}_{d})}{{F}^{2}}
\frac{2m_{K}^{2}-m_{\pi}^{2}}{M_{S^{\prime}}^{2}}
+\frac{4c^{\prime}_{m}}{F^{2}}\frac{(c^{\prime}_{m}-c^{\prime}_{d})2m_{\pi}^{2}+
c^{\prime}_{d}\left(s+m_{\pi}^{2}-m_{\eta^{(\prime)}}^{2}\right)}{M_{S^{\prime}}^{2}-s}\right]}\ ,
\end{array}
\label{formfactorRChT2}
\ee
where $S$ and $S^{\prime}$ correspond to the $a_{0}(980)$ and $a_{0}(1450)$ resonances, respectively.
The short-distance requirement that the form factors go to zero for $s\to\infty$ then implies the constraints
\cite{Jamin:2001zq}:
\be
4c_{d}c_{m}+4c^{\prime}_{m}c^{\prime}_{d}=F^{2}\ ,
\qquad
\frac{c_{m}}{M_{S}^{2}}(c_{m}-c_{d})+\frac{c^{\prime}_{m}}{M_{S^{\prime}}^{2}}(c^{\prime}_{m}-c^{\prime}_{d})=0\ .
\label{couplingrelation}
\ee
Not so much is known on the exact values of the couplings $c^{\prime}_{d,m}$ (and, to some extent, on $c_{d,m}$).
The estimate with only one scalar resonance led to $c_{d}=c_{m}$
and thus it seems plausible to keep this constraint in the case of two resonances.
One immediate consequence of the constraint and the second relation in Eq.~(\ref{couplingrelation}) is
$c^{\prime}_{d}=c^{\prime}_{m}$.
Then, the SFFs can be expressed, with $c_{m}$ and $c^{\prime}_{m}$ fulfilling $c_{m}^2+c^{\prime 2}_{m}=F^2/4$, as
\be
\begin{array}{l}
F_{0}^{\pi^{-}\eta^{(\prime)}}(s)=
c_{0}^{\pi^{-}\eta^{(\prime)}}\left[1+
\displaystyle{
\frac{4}{F^{2}}\left(\frac{c_{m}^2}{M_{S}^{2}-s}+\frac{c^{\prime 2}_{m}}{M_{S^{\prime}}^{2}-s}\right)}
\left(s+m_{\pi}^{2}-m_{\eta^{(\prime)}}^{2}\right)\right]\\[4ex]
\quad\longrightarrow
\displaystyle{\frac{c_{0}^{\pi^{-}\eta^{(\prime)}}}
{\left(M_{S}^{2}-s-iM_{S}\Gamma_{S}(s)\right)\left(M_{S^{\prime}}^{2}-s-iM_{S^{\prime}}\Gamma_{S^{\prime}}(s)\right)}}
\left\{\left(M_{S}^{2}-s\right)\left(M_{S^{\prime}}^{2}-s\right)\right.\\[4ex]
\qquad\quad\left.
+\displaystyle{\frac{4}{{F}^{2}}}
\left[c_{m}^{2}\left(M_{S^{\prime}}^{2}-s\right)+c^{\prime 2}_{m}\left(M_{S}^{2}-s\right)\right]
\left(s+m_{\pi}^{2}-m_{\eta^{(\prime)}}^{2}\right)\right\}\ ,
\end{array}
\label{B_W2}
\ee
once the energy-dependent widths have been incorporated into the scalar propagators.
Regarding the numerical values, we employ $c_{m}=41.9$ MeV \cite{Guo:2012yt} for the scalar coupling,
and $M_{S^{\prime}}=(1474\pm 19)$ MeV and $\Gamma_{S^{\prime}}=(265\pm 13)$ MeV \cite{pdg}
for the $a_{0}(1450)$ mass and width,
%(with a pole in the complex plane located at $\sqrt{s_{p}}=(1457-i161/2)$ MeV),
respectively.
In Fig.~\ref{BWplot2}, the normalized $\pi\eta^{(\prime)}$ SFFs obtained from Eq.~(\ref{B_W2})
in the approximation of considering two resonances are shown and compared with the single-resonance case.
Notice now that the normalized expressions depend on the mode.
While in the $\pi\eta$ case, one clearly sees a dominant peak corresponding to the $a_{0}(980)$
followed by a second smaller one in association with the $a_{0}(1450)$,
in the $\pi\eta^{(\prime)}$ case, two similar peaks located around both resonances are found.

\subsection{Elastic dispersion relation: Omn\`{e}s integral}
\label{elasticdispersionrelation}
A two-meson form factor is an analytic function in the whole complex plane except for the branch cut originated
as soon as the energy reaches the threshold for producing the first intermediate state where an imaginary part
is then developed.
The case in which the intermediate state is exactly the same as the final one is known as elastic
and the corresponding cut is called the unitary or elastic cut.
For the case at hand this cut starts at $s_{\rm th}=(m_{\pi^{-}}+m_{\eta^{(\prime)}})^{2}$
and the corresponding (elastic) unitarity relation for the scalar form factor reads
\be
{\rm{Im}}F_{0}^{\pi^{-}\eta^{(\prime)}}(s)=
\sigma_{\pi^{-}\eta^{(\prime)}}(s)F_{0}^{\pi^{-}\eta^{(\prime)}}(s)t_{10}^{\pi^{-}\eta^{(\prime)}*}(s)\ ,
\label{unitarityrelation}
\ee
where $t_{10}^{\pi^{-}\eta^{(\prime)}}(s)$ is the unitarized elastic $\pi^{-}\eta^{(\prime)}$ partial-wave,
with $I=1$ and $J=0$, of the scattering amplitude to be discussed later.
Analyticity, which relates the real and imaginary parts of the form factor,
is ensured through the use of a dispersion relation whose solution leads to the well-known
Omn\`{e}s integral \cite{Omnes:1958hv}.
When one subtraction is performed, the SFF can be written as\footnote{The
dispersive integral is uniquely specified up to a polynomial ambiguity.
This ambiguity is canceled by the subtraction function \cite{Pallante:2000hk}.
Both can be fixed from theory, for instance ChPT or RChT, or experimental data.
If the form factor is ``well-behaved'' at high-energies,
that is $\lim_{s\to\infty}F_{0}(s)=0$, the subtraction function can be fixed to a constant.}
\be
F_{0}^{\pi^{-}\eta^{(\prime)}}(s)=
F_{0}^{\pi^{-}\eta^{(\prime)}}(0)\exp\left[\frac{s}{\pi}\int_{s_{\rm th}}^{\infty}ds^{\prime}
\frac{\delta_{10}^{{\pi^{-}\eta^{(\prime)}}}(s^{\prime})}{s^{\prime}(s^{\prime}-s-i\varepsilon)}\right]\ ,
\label{Omnes}
\ee
where $\delta_{10}^{{\pi^{-}\eta^{(\prime)}}}$ is the phase shift associated to $t_{10}^{\pi^{-}\eta^{(\prime)}}$
and the value of the SFF at the origin has been chosen for convenience as the subtraction constant
(the subtraction point is then set to zero).

The so-called dispersive representation has been wide and successfully employed to describe
lots of phenomena and in particular data on exclusive hadronic tau decays
\cite{Boito:2008fq,Boito:2010me,Dumm:2013zh,Escribano:2013bca,Escribano:2014pya,Escribano:2014joa,
Gonzalez-Solis:2015nfa,Celis:2013xja,Bernard:2011ae,Bernard:2013jxa}\footnote{See also
Ref.~\cite{Kang:2013jaa} for the interesting case of $B_{l4}$ decays.}.
Unfortunately, for the $\pi^{-}\eta^{(\prime)}$ decay modes there is a lack of experimental data
either on the phase shifts or the decays spectra.
However, in the elastic region Watson's theorem \cite{Watson:1954uc} states that
the form factor phase equals that of the corresponding elastic scattering amplitude.
Thus, we can access this phase through the identification
\be
\phi_{\pi^{-}\eta^{(\prime)}}(s)\equiv\delta_{10}^{\pi^{-}\eta^{(\prime)}}(s)=
\arctan\frac{{\rm{Im}}t^{\pi^{-}\eta^{(\prime)}}_{10}(s)}{{\rm{Re}}t^{\pi^{-}\eta^{(\prime)}}_{10}(s)}\ .
\ee
Regarding the scattering amplitudes $\pi^{-}\eta\to\pi^{-}\eta$ and $\pi^{-}\eta^{\prime}\to\pi^{-}\eta^{\prime}$,
we have considered convenient here to use the expressions obtained within the global analysis of the
$U(3)\otimes U(3)$ one-loop meson-meson scattering amplitudes in ChPT including resonances, 
carried out in Ref.~\cite{Guo:2011pa}.
In that work, the partial-wave amplitudes have been properly deduced and unitarised through the $N/D$ method
\cite{Oller:1998zr,Oller:2000ma},
whose general simplified perturbative solution reads
\be
t_{IJ}^{PQ}(s)=\frac{\sigma_{PQ}(s)N^{PQ}_{IJ}(s)}{1+g_{PQ}(s)N^{PQ}_{IJ}(s)}\ ,
\label{NDmethod}
\ee
and finally applied to fit the available scattering amplitudes' phase shifts.
In Eq.~(\ref{NDmethod}), $PQ$ refers to the interacting meson-meson system in question,
$g_{PQ}(s)$ are the dimeson one-loop scalar functions defined in Eq.~(33) of Ref.~\cite{Guo:2011pa} and 
$N^{PQ}_{IJ}(s)$ contain the expressions of the partial-wave amplitudes up to next-to-leading order.

\begin{figure}
\centering
\includegraphics[scale=0.45]{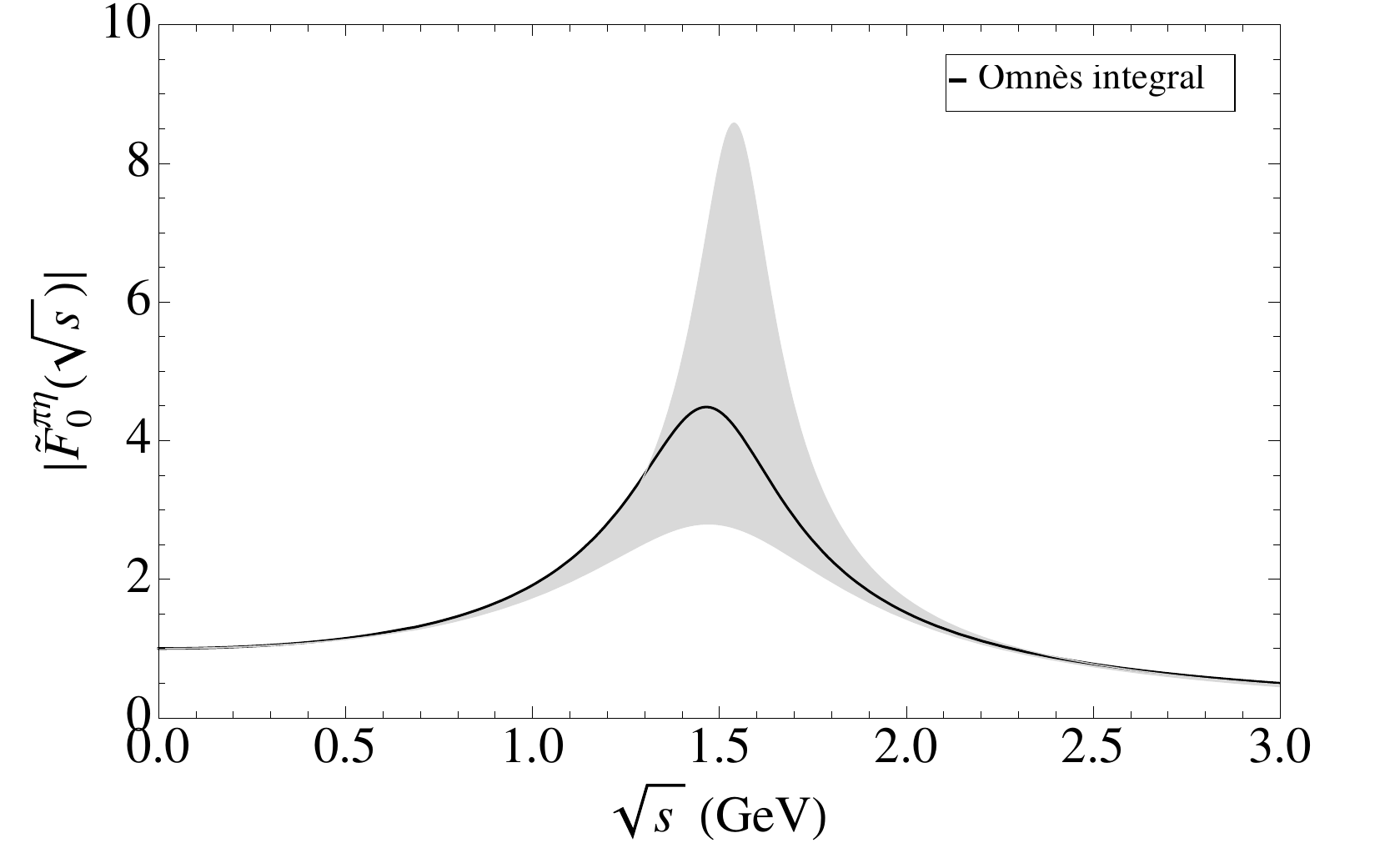}
\includegraphics[scale=0.45]{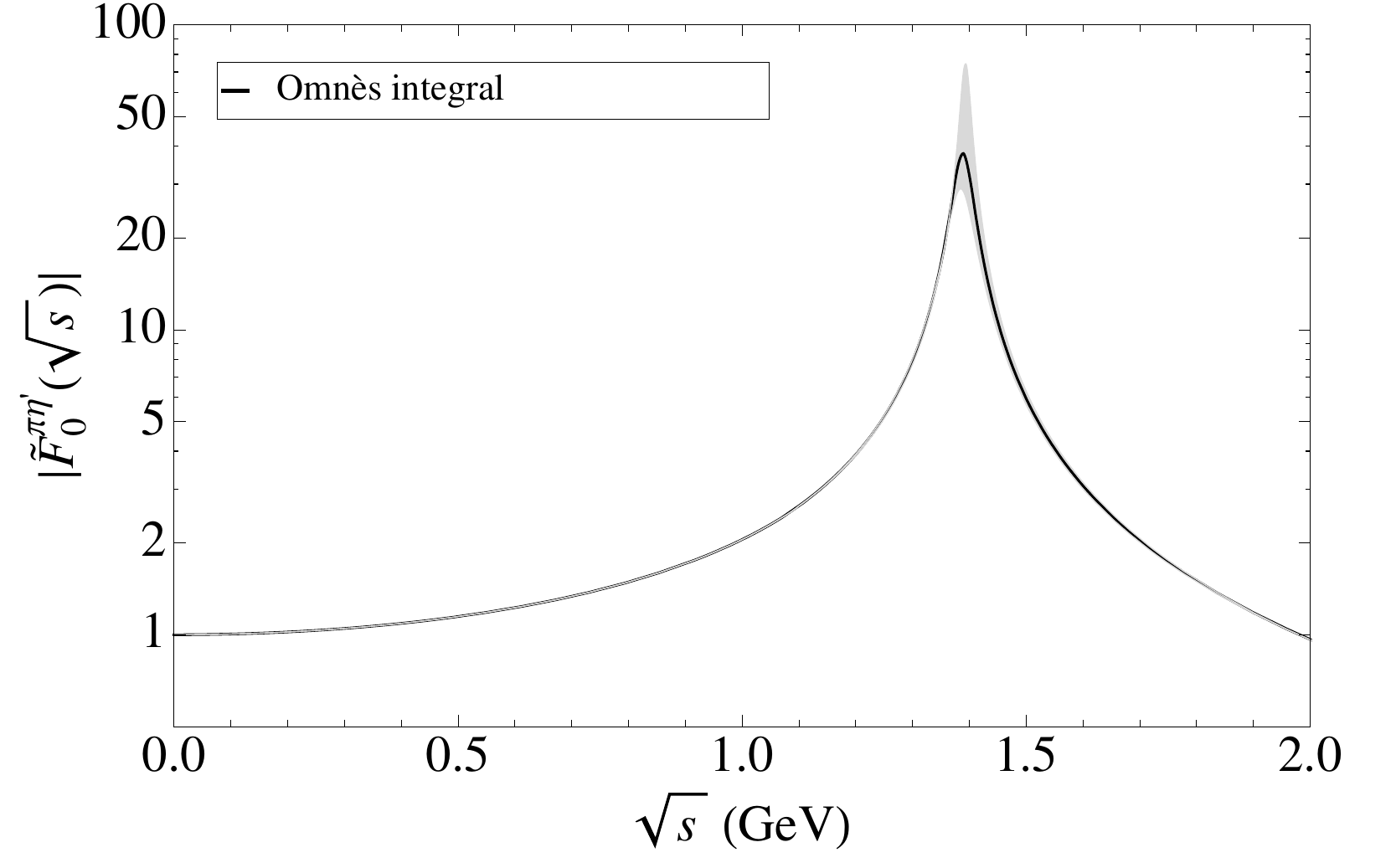}
\caption{Normalized
$\pi^{-}\eta$ (left plot) and $\pi^{-}\eta^{\prime}$ (right plot) scalar form factors
as obtained from the elastic dispersion relation (Omn\`es integral) described in Sec.~\ref{elasticdispersionrelation}.
The grey error bands account for the (uncorrelated) uncertainty on the input values from Ref.~\cite{Guo:2012yt}.}
\label{elastic}
\end{figure}

In Fig.~\ref{elastic}, we represent the elastic SFFs obtained using Eq.~(\ref{Omnes})
and the results from the updated analysis of Ref.~\cite{Guo:2012yt}
as the input values of the theory: couplings, masses, etc.
Specifically, we are using the values in Eq.~(45) of this reference,
neglecting error correlations since we ignore them.
For the normalizations, as stated,
we have chosen $F_{0}^{\pi^{-}\eta}(0)=0.92$ and $F_{0}^{\pi^{-}\eta^{\prime}}(0)=0.05$
from Eq.~(\ref{SFF0})\footnote{These
inputs could be checked with lattice QCD simulations incorporating isospin-breaking corrections.}.
The plots show a resonant region at around $1.4$ GeV
which may be attributed to the effect of the $a_{0}(1450)$ resonance.
This presence and the absence of a corresponding peak for the $a_{0}(980)$ 
is explained because the former resonance appears in the $s$-channel of the scattering amplitude
while the latter only in the crossed $t$ and $u$ channels.

It can be verified that these SFFs can also be written in a closed expression as 
\cite{Jamin:2001zq,Babelon:1976kv,Spearman}
\be
F_{0}^{\pi^{-}\eta^{(\prime)}}(s)=
\prod_{j}\frac{1}{\left(1-s/s_{z_j}\right)}\frac{F_{0}^{\pi^{-}\eta^{(\prime)}}(0)}
{\left(1+g_{\pi^{-}\eta^{(\prime)}}(s)N^{\pi^{-}\eta^{(\prime)}}_{10}(s)\right)}\ .
\label{elasticclosed}
\ee
The $s_{z_j}$ are the locations of the zeros of the inverse of the denominator functions,
$D^{\pi^{-}\eta^{(\prime)}}(s)\equiv 1+g_{\pi^{-}\eta^{(\prime)}}(s)N^{\pi^{-}\eta^{(\prime)}}_{10}(s)$,
which have to be removed in the form factors.
In our specific case, the zero is placed at $s_{z_1}=1.9516$ GeV$^2$ corresponding to
the bare (squared) mass of the scalar octet $S_{8}$ \cite{Guo:2012yt}.
As a consistency check, we have verified that the results obtained with Eq.~(\ref{Omnes})
are reproduced using the closed expression in Eq.~(\ref{elasticclosed}). 
Inspired by the works of Refs.~\cite{Iwamura:1976fc,Babelon:1976kv,Iwamura:1977ds,
Kamal:1979be,Kamal:1980mw,Kamal:1987nm,Kamal:1988ub,Oller:2000ug,
Bjorken:1960zz,Lee:1974pm,Sorensen:1981vu,Basdevant:1978tx,Basdevant:1978wb},
we propose to obtain the analogous expression of Eq.~(\ref{elasticclosed})
valid for the description of the coupled-channels case.
In this respect, our closed form solution for this case giving the participant SFFs
appears numerically advantageous
(instead of the more common iterative solution of the coupled integro-differential set of equations)
for the Monte Carlo event generator performance \cite{Actis:2010gg},
specially if our expressions are to be used for fitting the resonance parameters appearing in the 
form factors.
The method is detailed in appendix \ref{App}. 

\subsection{Two coupled channels}
\label{twocoupledchannels}
We first consider the two coupled channels case involving the $\pi^{-}\eta$ and $\pi^{-}\eta^{\prime}$ cuts.
The two-meson loop function and the required partial-wave scattering amplitudes are organized in symmetric matrices
given by
\be
g(s)=
\begin{pmatrix}
             g_{\pi^{-}\eta} &0\\
             0 & g_{\pi^{-}\eta^{\prime}}
\end{pmatrix}
\ ,
\qquad
N_{10}(s)=
\begin{pmatrix}
             N_{\pi^{-}\eta\to\pi^{-}\eta} & N_{\pi^{-}\eta\to\pi^{-}\eta^{\prime}}\\
             N_{\pi^{-}\eta^{\prime}\to\pi^{-}\eta} & N_{\pi^{-}\eta^{\prime}\to\pi^{-}\eta^{\prime}}
\end{pmatrix}
\ ,
\ee
where each entry of the matrix $N(s)$ (omitting the $IJ$ quantum numbers) reads
$N_{ij}(s)=T^{\mathcal{O}(p^{4})}_{ij}(s)-g_{i}(s)\left(T_{ij}^{\mathcal{O}(p^{2})}(s)\right)^{2}$,
for $i,j=1,2$, with $T^{\mathcal{O}(p^{4})}_{ij}(s)$ referring to the corresponding partial-wave amplitude
at $\mathcal{O}(p^{4})$, which includes the $\mathcal{O}(p^{2})$ term,
the $\mathcal{O}(p^{4})$ contributions arising from wave-function renormalisation of the fields and, finally,
the explicit $\mathcal{O}(p^{4})$ resonance-exchange and one-loop diagrams in the $s$-channel
as well as in the crossed $t$ and $u$ channels.
Written in this way, the double counting of loop contributions in the $s$-channel is avoided.
For the sake of clarity, the Eq.~(\ref{coupledchannels}) in appendix \ref{App}
applied to this particular case would read
\be
\begin{array}{l}
\begin{pmatrix}
	F_{0}^{\pi^{-}\eta}(s)\\
	F_{0}^{\pi^{-}\eta^{\prime}}(s)
\end{pmatrix}
=
\displaystyle
\frac{1}{{\rm Det}[D_{IJ}(s)]}\times\\[3ex]
\begin{pmatrix}
	1+g_{\pi^{-}\eta^{\prime}}(s)N_{\pi^{-}\eta^{\prime}\to\pi^{-}\eta^{\prime}}(s)&
	-g_{\pi^{-}\eta}(s)N_{\pi^{-}\eta\to\pi^{-}\eta^{\prime}}(s)\\
	-g_{\pi^{-}\eta^{\prime}}(s)N_{\pi^{-}\eta^{\prime}\to\pi^{-}\eta}(s)&
	1+g_{\pi^{-}\eta}(s)N_{\pi^{-}\eta\to\pi^{-}\eta}(s)
\end{pmatrix}
%\begin{pmatrix}
%	1&0\\
%	0&1
%\end{pmatrix}
\begin{pmatrix}
	F_{0}^{\pi^{-}\eta}(0)\\
	F_{0}^{\pi^{-}\eta^{\prime}}(0)
\end{pmatrix}
\ ,
\end{array}
\label{2ccpietatopietap}
\ee
where the subtraction point $s_{0}$ has been set to zero in analogy with
Refs.~\cite{Boito:2008fq,Escribano:2013bca,Escribano:2014joa}.
The determinant of the matrix $D_{IJ}(s)$, defined in Eq.~(\ref{DIJmatrix}) of appendix \ref{App},
may vanish for some values of $s$.
To get rid of these possible zeros we factorize them in the same way
it has been done in Eq.~(\ref{elasticclosed}) for the single-channel case.
These singularities can be understood as dynamically generated resonances
appearing after the rescattering of the pseudoscalars mesons involved.
In our case, ${\rm Det}[D_{IJ}(s)]$ is seen to vanish again at $s=1.9516$ GeV$^{2}$
for the same reason given in the elastic case.

\begin{figure}
\centering
\includegraphics[scale=0.4]{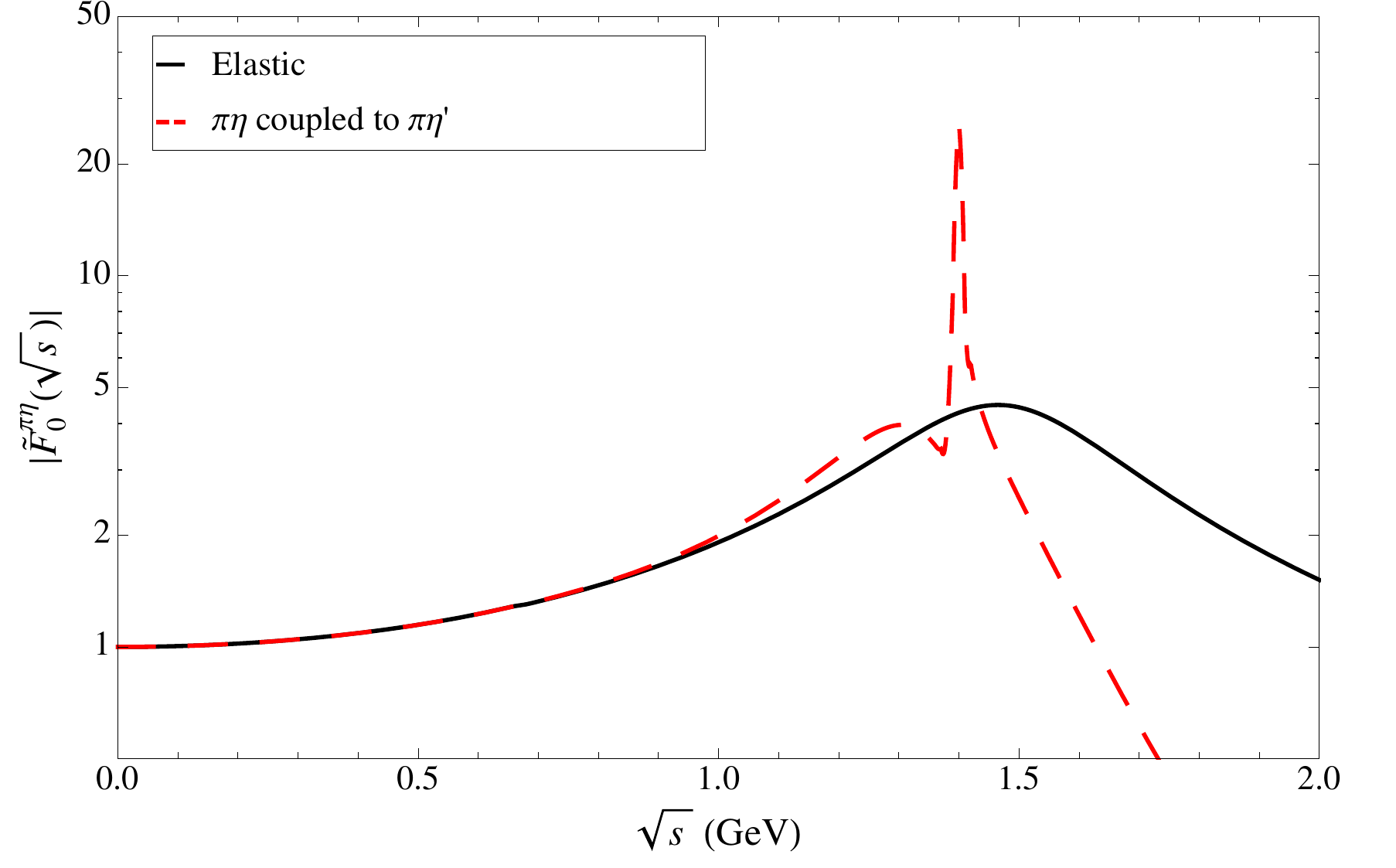}
\includegraphics[scale=0.4]{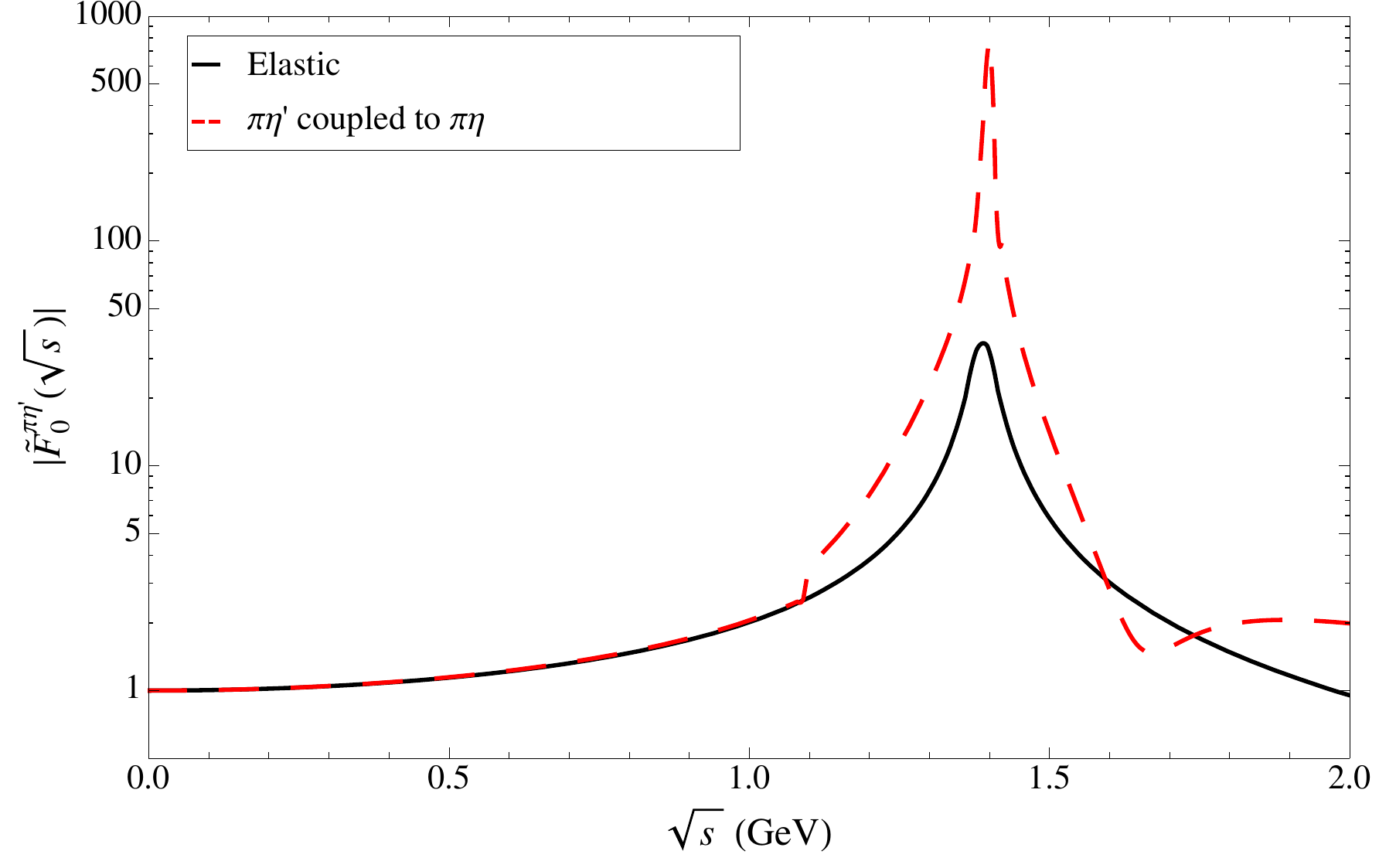}
\caption{Normalized
$\pi^{-}\eta$ (left plot) and $\pi^{-}\eta^{\prime}$ (right plot) scalar form factors
as obtained from Eq.~(\ref{2ccpietatopietap}) in Sec.~\ref{twocoupledchannels} (red dashed curves)
compared to the corresponding elastic cases (black solid curves).}
\label{cc1}
\end{figure}

In Fig.~\ref{cc1}, we display the results obtained from Eq.~(\ref{2ccpietatopietap}).
The $\pi^{-}\eta$ SFF coupled to $\pi^{-}\eta^{\prime}$ (left plot)
and \emph{vice versa} (right plot) are compared to their respective elastic case.
As seen, the $\pi^{-}\eta$ SFF develops a thin peak at around $1.4$ GeV followed by a hard drop.
We can also observe that, generically, the neat effect of coupling the $\pi^{-}\eta^{\prime}$ channel
into the $\pi^{-}\eta$ SFF is small.
On the contrary, the impact of the $\pi^{-}\eta$ channel in the description of the $\pi^{-}\eta^{\prime}$ SFF is large,
the resonance region is highly enhanced.
Needless to say, the coupled-channels effects start at the $\pi^{-}\eta$ and $\pi^{-}\eta^{\prime}$ thresholds,
respectively, and in case these inelasticities were switched off the elastic description would be recovered. 

\begin{figure}
\centering
\includegraphics[scale=0.4]{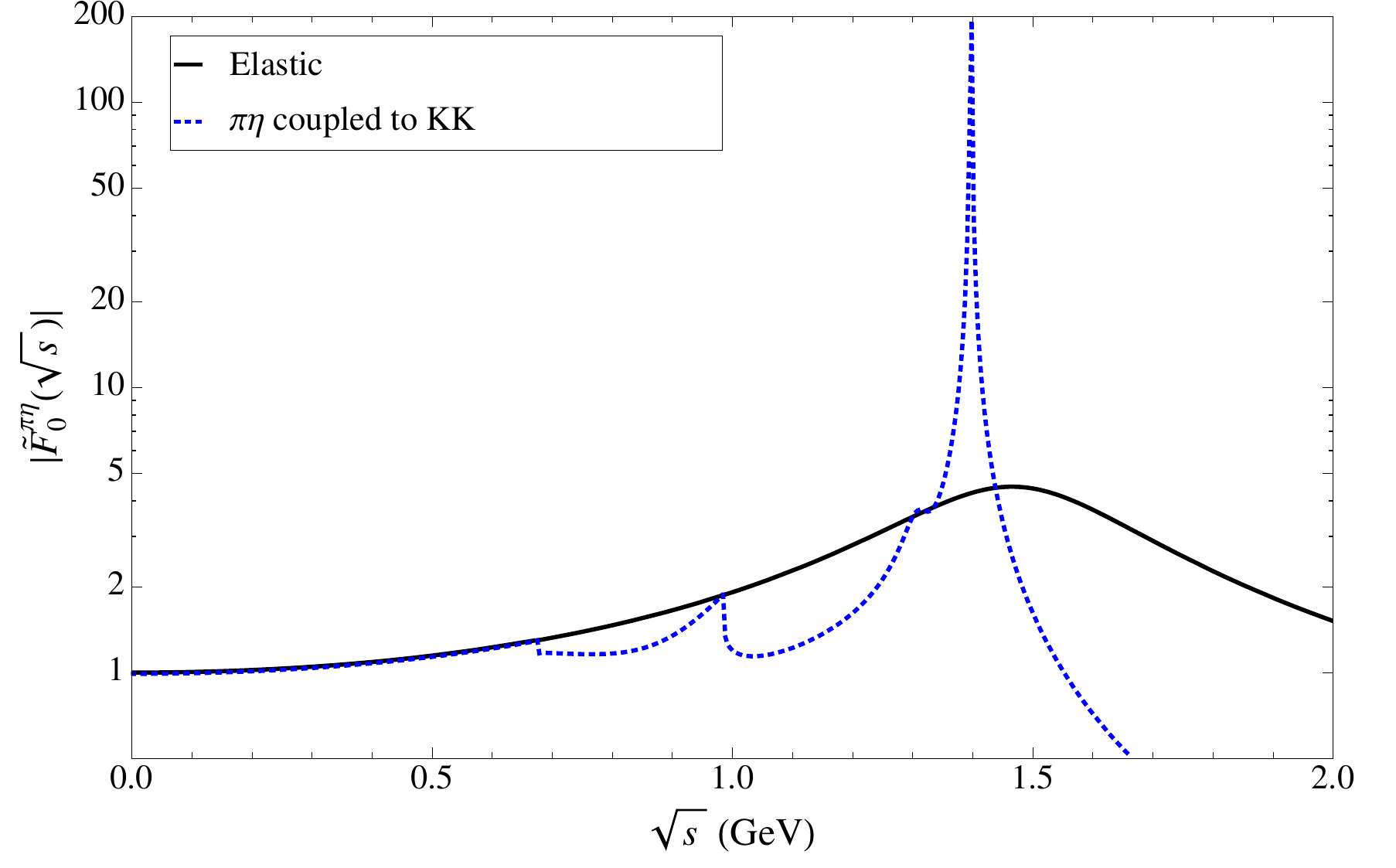}
\includegraphics[scale=0.4]{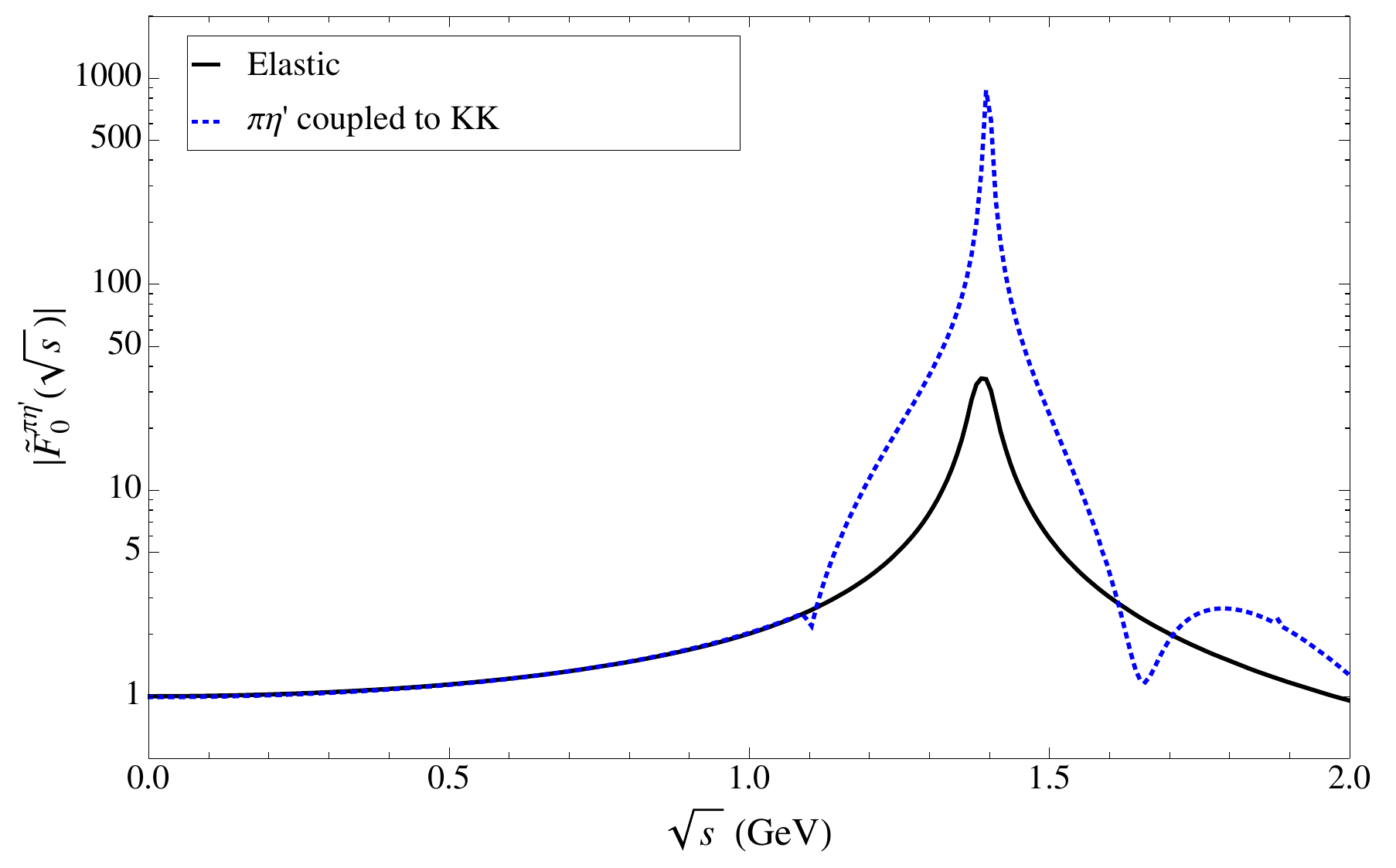}
\caption{Normalized
$\pi^{-}\eta$ (left plot) and $\pi^{-}\eta^{\prime}$ (right plot) scalar form factors
as obtained from Eq.~(\ref{coupledchannels}) in App.~\ref{App} (blue dotted curves)
compared to the corresponding elastic cases (black solid curves).}
\label{cc2}
\end{figure}

Analogously, we can consider the $K^{-}K^{0}$ cut which is located between
the $\pi^{-}\eta$ and $\pi^{-}\eta^{\prime}$ thresholds.
A \emph{priori}, one would expect the intermediate $\bar{u}d$-like scalar to strongly couple to the $K^{-}K^{0}$ system
\cite{Astier:1967zz}.
We emphasize that the value at the origin of the $K^{-}K^{0}$ SFF,
as computed from RChT in a similar way to Eq.~(\ref{B_W}) for the $\pi^{-}\eta^{(\prime)}$ ones,
is $F_{0}^{K^{-}K^{0}}(0)=1$
(this can be easily understood observing that the kaon mass difference is very small compared to
the chiral symmetry breaking scale),
and therefore its weight may be relevant.
This is corroborated in Fig.~\ref{cc2}, where the $\pi^{-}\eta^{(\prime)}$ SFFs coupled to $K^{-}K^{0}$ are shown.
Notice that this time the effect on the $\pi^{-}\eta$ SFF is sizable.
After a small dip at the $\pi^{-}\eta$ threshold one can see a small peak at the $K^{-}K^{0}$ threshold and
a significant enhancement between $1.3$ and $1.45$ GeV with respect to the elastic case.
This is one interesting result which may help to unveil the somewhat ``exotic'' nature of the scalar resonances
that couple to the $\bar{u}d$ operator.
Suggestions like a tetraquark interpretation as well as molecular $K\bar{K}$ threshold states
exist in the literature\footnote{See
{\textit{e.g.}} the ``Note on scalar mesons below 2 GeV'' in Ref.~\cite{pdg}.}.

\subsection{Three coupled channels}
\label{threecoupledchannels}
\begin{figure}
\centering
\includegraphics[scale=0.4]{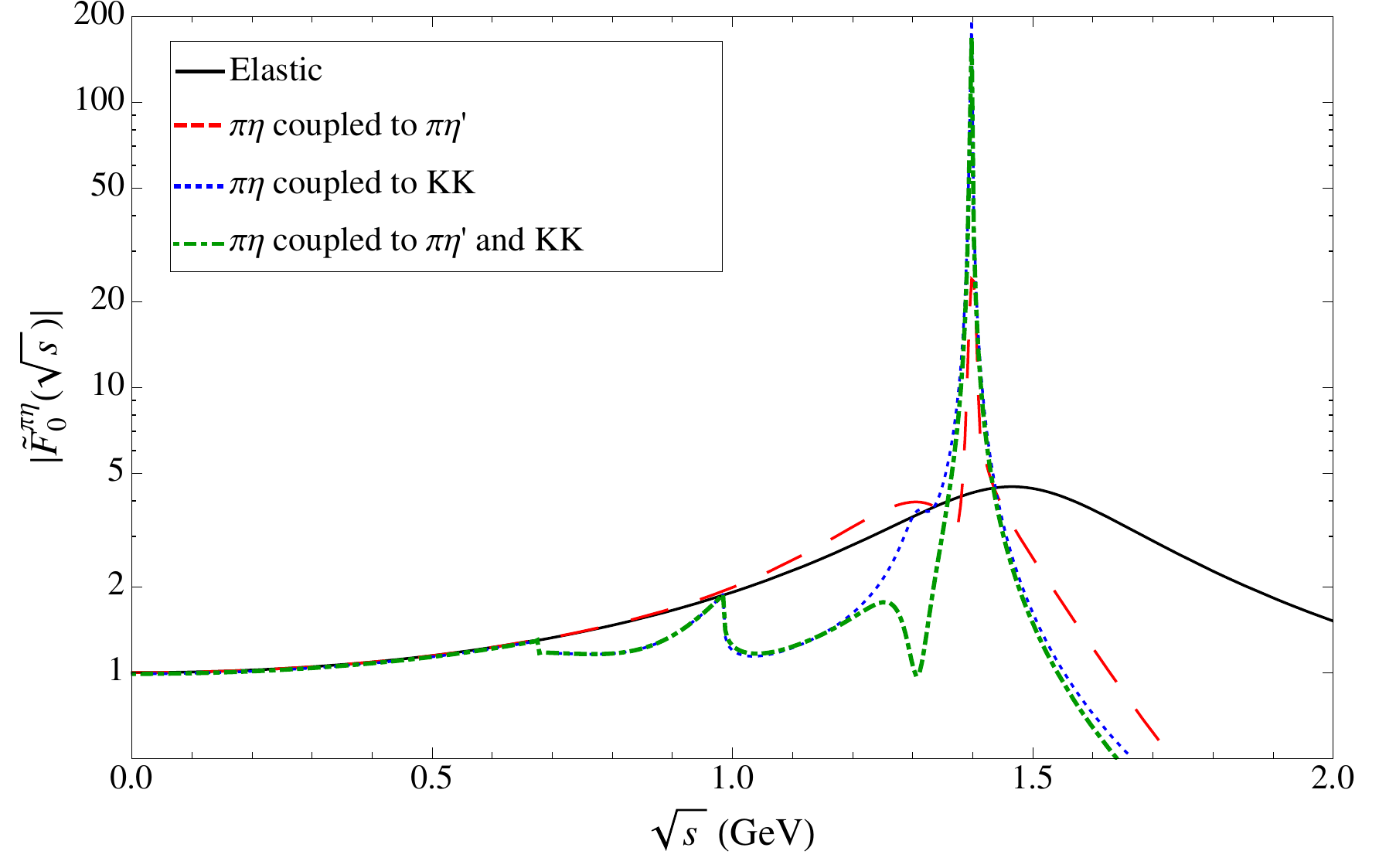}
\includegraphics[scale=0.4]{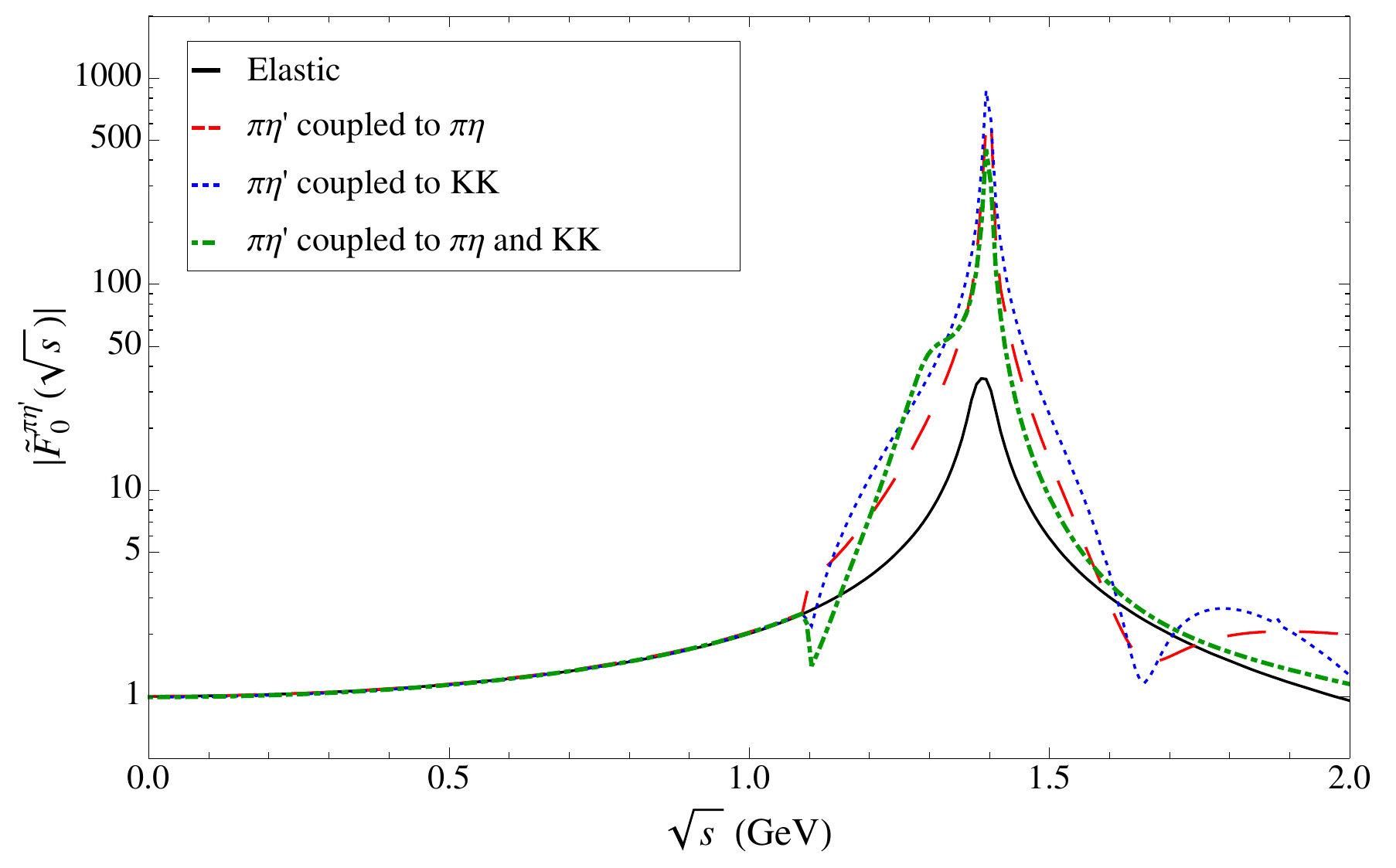}
\caption{Normalized
$\pi^{-}\eta$ (left plot) and $\pi^{-}\eta^{\prime}$ (right plot) scalar form factors
as obtained from the three coupled channels discussion in Sec.~\ref{threecoupledchannels}
(green dot-dashed curve)
compared to the corresponding elastic (black solid curves)
and two coupled channels (red dashed and blue dotted curves) cases.}
\label{cc3}
\end{figure}

\begin{figure}
\centering
\includegraphics[scale=0.7]{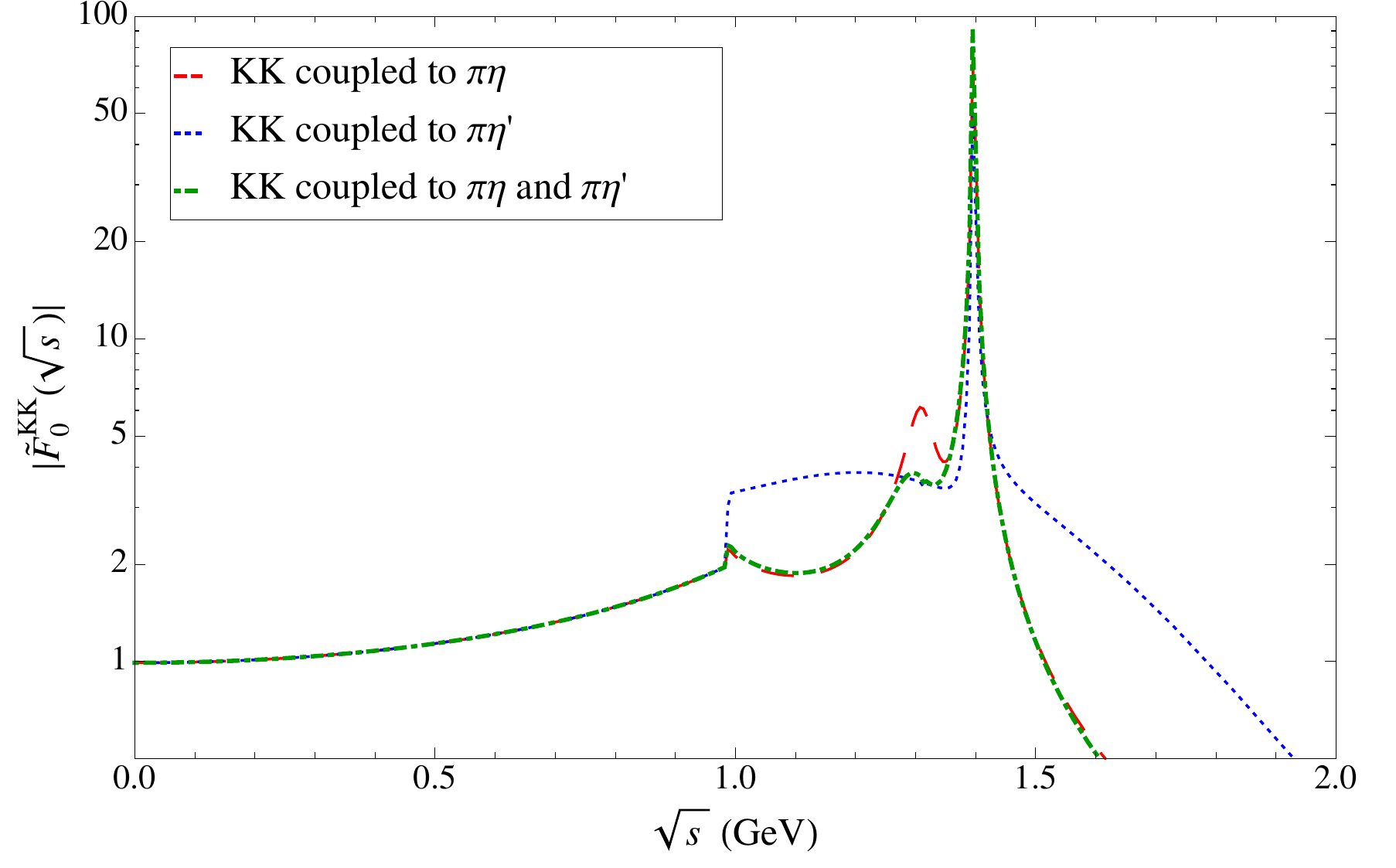}
\caption{Normalized $K^{-}K^{0}$ scalar form factor
as obtained from the three coupled channels discussion in Sec.~\ref{threecoupledchannels}
(green dot-dashed curve)
compared to the corresponding two coupled channels (red dashed and blue dotted curves) cases.}
\label{cc3KK}
\end{figure}

Let us now turn to the case in which the $\pi^{-}\eta$, $\pi^{-}\eta^{\prime}$ and $K^{-}K^{0}$ cuts
are considered simultaneously\footnote{The
$\pi^-\pi^0$ cut is safely neglected because no resonance contributions to this channel are allowed
at first order in isospin breaking.
However, its low-energy limit has been derived in Ref.~\cite{Descotes-Genon:2014tla} 
in a model-independent way because of its importance in producing a sizable CP-violating asymmetry
in the dipion tau decays, albeit only very close to the $\pi\pi$ threshold.}.
This requires to perform a calculation when the three channels are coupled to each other.
In this case, the matrices encoding the corresponding scalar loop functions and partial-wave amplitudes
are given by
\be
g(s)=\begin{pmatrix}
             g_{\pi^{-}\eta} & 0&0\\
             0&g_{\pi^{-}\eta^{\prime}}&0\\
             0&0 & g_{K^{-}K^{0}}
\end{pmatrix}
\ ,
\ee
and
\be
N_{10}(s)=
\begin{pmatrix}
             N_{\pi^{-}\eta\to\pi^{-}\eta} & N_{\pi^{-}\eta\to\pi^{-}\eta^{\prime}} & N_{\pi^{-}\eta\to K^{-}K^{0}}\\
             N_{\pi^{-}\eta^{\prime}\to\pi^{-}\eta} & N_{\pi^{-}\eta^{\prime}\to\pi^{-}\eta^{\prime}} &
             N_{\pi^{-}\eta^{\prime}\to K^{-}K^{0}}\\
             N_{K^{-}K^{0}\to\pi^{-}\eta} & N_{K^{-}K^{0}\to\pi^{-}\eta^{\prime}} & N_{K^{-}K^{0}\to K^{-}K^{0}}
\end{pmatrix}
\ ,
\ee
respectively.
From the analogous expression to Eq.~(\ref{2ccpietatopietap}) for the case of three coupled channels,
which we do not quote explicitly, we obtain
the $\pi^{-}\eta$ SFF coupled to $\pi^{-}\eta^{\prime}$ and $K^{-}K^{0}$
as well as the $\pi^{-}\eta^{\prime}$ SFF coupled to $\pi^{-}\eta$ and $K^{-}K^{0}$.
In Fig.~\ref{cc3}, we provide a graphical account of these results compared with all previous cases. 
For the $\pi\eta$ SFF, the three coupled channels solution follows closely the one obtained coupling 
the $\pi\eta$ and $K^{-}K^{0}$ channels, except for the region between $1.2$ and $1.3$ GeV
where a dip appears first.
On the contrary, for $\pi^{-}\eta^{\prime}$ SFF, the three coupled channels solution does not appear to be
significantly dominated in the inelastic region by any of the two coupled channels results.
In addition, we get the $K^{-}K^{0}$ SFF coupled to the $\pi^{-}\eta^{(\prime)}$ systems as shown in Fig.~\ref{cc3KK}.
In this case, the three coupled channels solution resembles very much to the $\pi^{-}\eta$ channel
coupled to $K^{-}K^{0}$ apart from the region between $1.3$ to $1.4$ GeV
where the peak in the two-channels case almost disappears in the three-channels solution.
Finally, it is worth mentioning that the effects of the $\pi\eta^{(\prime)}\to\pi(\pi)\gamma$ channels 
should be considered as well.
However, the devoted discussion of these contributions in Ref.~\cite{Descotes-Genon:2014tla}
reveals that either subleading isospin-breaking effects of the $\rho$ contribution to the one-pion final state
or phase space considerations in the two-pion channel suppress these channels enough
so as to neglect them at the current level of uncertainty.

\section{Spectra and branching ratio predictions}
\label{predictions}
\begin{figure}
\centering
\includegraphics[scale=0.85]{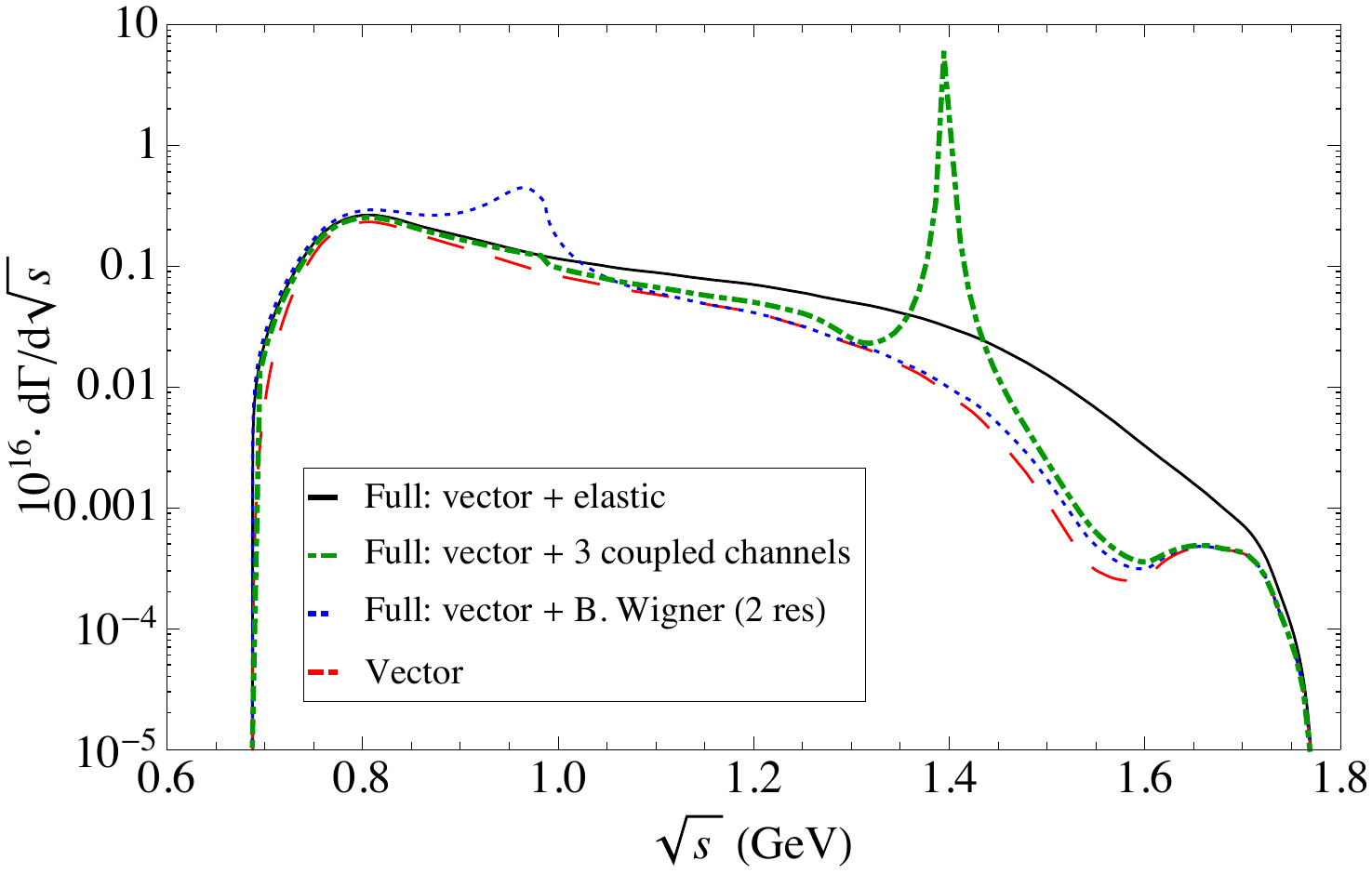}
\caption{Total differential decay width distribution for $\tau^{-}\to\pi^{-}\eta\nu_{\tau}$
as a function of the $\pi^-\eta$ invariant mass for the different parameterizations of the $\pi^-\eta$ SFF:
Breit-Wigner formula incorporating two resonances (blue dotted curve),
elastic dispersion relation (solid black curve),
and three coupled channels solution (green dot-dashed curve).
The vector contribution alone is also included for completeness (red dashed curve).}
\label{distributionpieta}
\end{figure}

\begin{figure}
\centering
\includegraphics[scale=0.8]{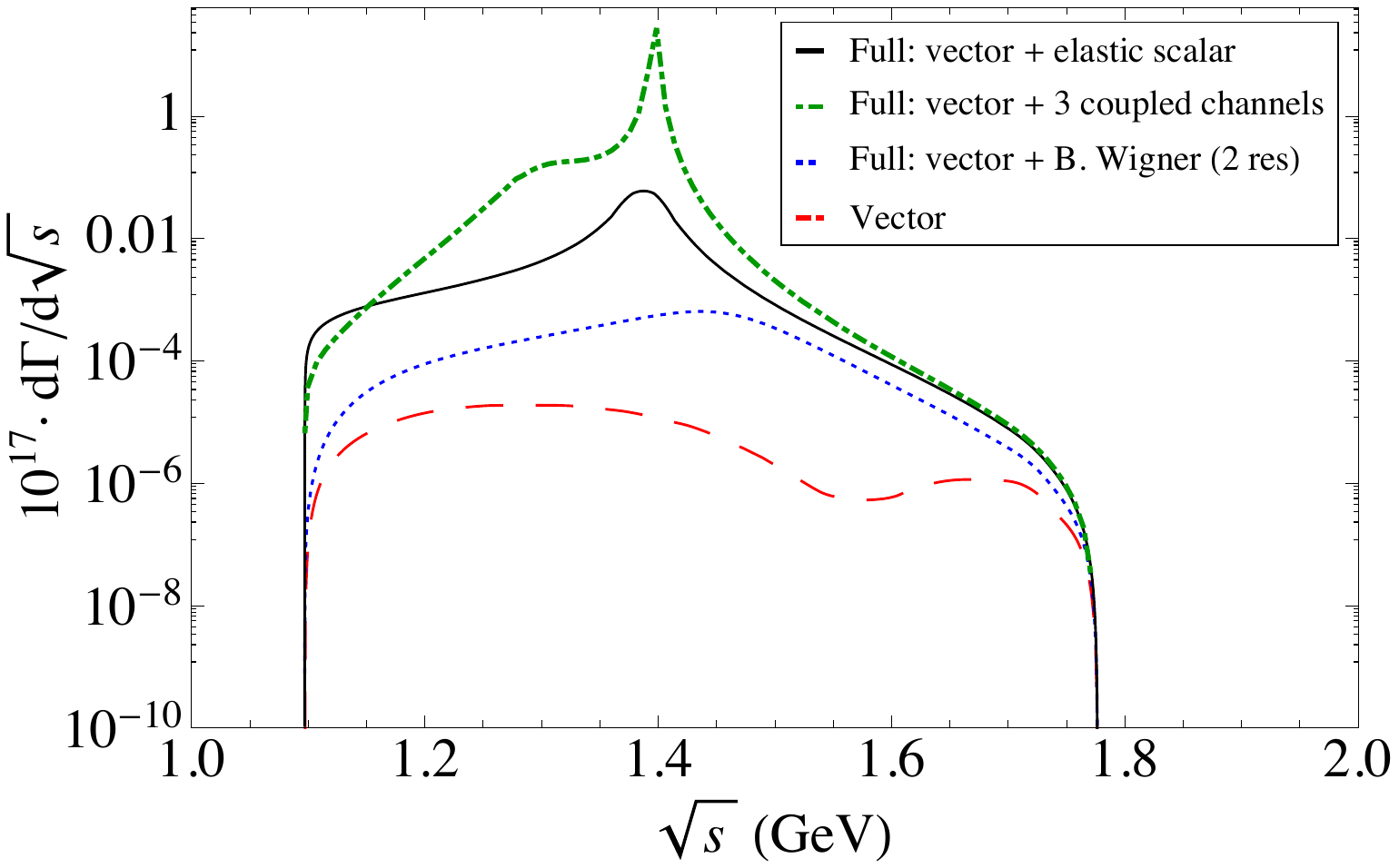}
\caption{Total differential decay width distribution for $\tau^{-}\to\pi^{-}\eta^{\prime}\nu_{\tau}$
as a function of the $\pi^-\eta^{\prime}$ invariant mass for the different parameterizations of the $\pi^-\eta^{\prime}$ SFF:
Breit-Wigner formula incorporating two resonances (blue dotted curve),
elastic dispersion relation (solid black curve),
and three coupled channels solution (green dot-dashed curve).
The vector contribution alone is also included for completeness (red dashed curve).}
\label{distributionpietap}
\end{figure}

The vector and scalar form factors as described in Secs.~\ref{VectorFormFactor} and \ref{ScalarFormFactors}
finally enter Eq.~(\ref{width}) to predict the partial width of the decays $\tau^{-}\to\pi^{-}\eta^{(\prime)}\nu_{\tau}$.
The corresponding invariant mass distributions (decay spectra) are plotted in
Figs.~\ref{distributionpieta} and \ref{distributionpietap}, respectively,
and the predicted branching ratios are given and compared to other authors' results in
Tables~\ref{BRpieta} and \ref{BRpietap}.
In the following, we discuss the two reactions separately.

\subsection{$\tau^{-}\to\pi^{-}\eta\nu_{\tau}$}
In Fig.~\ref{distributionpieta}, we display the total differential decay width distribution of $\tau^{-}\to\pi^{-}\eta\nu_{\tau}$
as a function of the $\pi^-\eta$ invariant mass for the different parameterizations of the $\pi^-\eta$ SFF
discussed in the text, that is, the Breit-Wigner formula incorporating two resonances (blue dotted curve),
the elastic dispersion relation (solid black curve),
and the three coupled channels solution (green dot-dashed curve).
For completeness, the vector contribution alone is also included (red dashed curve).
As seen, the low-energy part of the spectrum, ranging from the $\pi^{-}\eta$ threshold to 1.2 GeV,
is mainly dominated by the vector contribution associated to the $\rho(770)$ resonance.
Only in the case of the Breit-Wigner description of the SFF the low-energy region is instead dominated by the
$a_{0}(980)$ scalar resonance whose manifestation is clearly visible around 1 GeV and then suppressed.
On the contrary, the scalar contribution as obtained either from the elastic dispersion relation
through the single channel Omn\`{e}s integral or the three coupled channels solution
dominates the energy region of the mass distribution above 1.2 GeV.
In the first case the distribution falls off smoothly,
while in the latter a sizable peak around $1.4$ GeV appears
due to the consideration of the $K^{-}K^{0}$ intermediate state
that could be attributed to the effect of the $a_{0}$(1450) resonance.
Finally, in the upper part of the spectrum, although suppressed,
the vector contributions from the $\rho^{\prime}$ and $\rho^{\prime\prime}$ resonances are suggested.

\begin{table}
\centering
\begin{tabular}{|l|l|l|l|}
\hline
BR$_V\times 10^{5}$ & BR$_S\times 10^{5}$ & BR$\times 10^{5}$ & Reference\\
\hline
$0.25$ & $1.60$ & $1.85$ & Tisserant, Truong \cite{Tisserant:1982fc}\\
$0.12$ & $1.38$ & $1.50$ & Bramon, Narison, Pich \cite{Bramon:1987zb}\\
$0.15$ & $1.06$ & $1.21$ & Neufeld, Rupertsberger \cite{Neufeld:1994eg}\\
$0.36$ & $1.00$ & $1.36$ & Nussinov, Soffer \cite{Nussinov:2008gx}\\
$[0.2,0.6]$ & $[0.2,2.3]$ & $[0.4,2.9]$ & Paver, Riazuddin \cite{Paver:2010mz}\\
$0.44$ & $0.04$ & $0.48$ & Volkov, Kostunin \cite{Volkov:2012be}\\
$0.13$ & $0.20$ & $0.33$ & Descotes-Genon, Moussallam \cite{Descotes-Genon:2014tla}\\
\hline\hline
BR$_V\times 10^{5}$ & BR$_S\times 10^{5}$ & BR$\times 10^{5}$ & Our analysis\\
\hline
$0.26\pm 0.02$ & $0.72^{+0.46}_{-0.22}$ & $0.98\pm0.51$ &Breit-Wigner [$a_{0}(980)$]\\
$0.26\pm 0.02$ & $0.48^{+0.29}_{-0.14}$ & $0.74\pm0.32$ &Breit-Wigner  [$a_{0}(980)+a_{0}(1450)$]\\
$0.26\pm 0.02$ & $0.10^{+0.02}_{-0.03}$ & $0.36\pm0.04$ & Elastic dispersion relation\\
$0.26\pm 0.02$ & $0.15\pm 0.09$	& $0.41\pm 0.09$	& 2 coupled channels ($\pi^{-}\eta$ \& $\pi^{-}\eta^{\prime}$)\\
$0.26\pm 0.02$ & $1.86\pm 0.11$	& $2.12\pm 0.11$	& 2 coupled channels ($\pi^{-}\eta$ \& $K^{-}K^{0}$)\\
$0.26\pm 0.02$ & $1.41\pm 0.09$	& $1.67\pm 0.09$	& 3 coupled channels\\
\hline\hline
& & BR$\times 10^{5}$ & Experimental collaboration\\
\hline
& & $<14$  ($95\%$ CL) & CLEO		\cite{Bartelt:1996iv}\\
& & $<7.3$ ($90\%$ CL) & Belle		\cite{Hayasaka:2009zz}\\
& & $<9.9$ ($95\%$ CL) & BaBar	\cite{delAmoSanchez:2010pc}\\
\hline
\end{tabular}
\caption{Vector, scalar and total contributions to the branching ratio (BR) of $\tau^{-}\to\pi^-\eta\nu_{\tau}$.
\emph{Upper part}:
results from previous phenomenological analyses.
\emph{Mid part}:
results from our analysis depending on the description of the $\pi^-\eta$ SFF used.
The source of uncertainty in the BRs arises from the errors on $\varepsilon_{\pi\eta}$
(the only source for the VFF and the SFF based on coupled channels)
and from the (uncorrelated) errors on the SFF input values.
\emph{Lower part}:
Current experimental upper bounds.}
\label{BRpieta}
\end{table}

In Table~\ref{BRpieta}, we present the results of our analysis for the integrated branching ratio
of $\tau^{-}\to\pi^{-}\eta\nu_{\tau}$
attending to the different parameterizations of the $\pi^-\eta$ SFF.
The values for the vector contribution, the scalar one, and the total branching ratio are shown separately.
Our results for each contribution are compared with previous phenomenological analyses existing in the literature.
The current experimental upper bounds are also included for comparison.
For the vector contribution, it is worth mentioning again that we benefit from the experimentally well-known
$\pi^{-}\pi^{0}$ VFF of $\tau^{-}\to\pi^{-}\pi^{0}\nu_{\tau}$ decays to fix this contribution 
(and the one to $\pi^-\eta^\prime$) up to a constant factor.
In this manner, the vector contribution to the $\pi^-\eta^{(\prime)}$ decays can be considered as model independent
since they are extracted directly from data.
To first order in isospin breaking, this constant factor is nothing else than the
$\pi^{0}$-$\eta^{(\prime)}$ mixing angle $\varepsilon_{\pi\eta^{(\prime)}}$, that is,
the normalization of the VFF at the origin.
This same normalization, see Eq.~(\ref{width}), appears as a global prefactor in the evaluation of the branching ratios.
Our predictions are pretty sensitive to the isospin-violating mixing angles $\varepsilon_{\pi\eta^{(\prime)}}$,
whose uncertainties become an important source of error.
Thus, precise determinations of these mixing angles would be very welcome.
A second important source of uncertainty is the intrinsic error associated to the SFF\footnote{This
error arises for the case of the Breit-Wigner formula from the (uncorrelated) errors of the resonance(s) parameters
and for the elastic dispersion relation from the (uncorrelated) errors of the input values from Ref.~\cite{Guo:2012yt}.
However, for the three coupled channels solution, we do not provide an error for the SFF
since the guess of using uncorrelated parameters produce large uncertainties resulting in predictions compatible with zero.
In line with this, it is pointed out in Ref.~\cite{Descotes-Genon:2014tla}
that an uncertainty smaller than 20\% in the $f_0^{\pi\eta}$ at 1 GeV would allow to improve the bounds
on a charged Higgs obtained from $B\to\tau\nu_\tau$ decays.
Our previous remark makes clear that this is not possible at present.}.
From the table, we observe that the obtained values for the vector contribution to the $\pi^-\eta$ VFF
are in line with existing ones.
The error stated comes from the $\varepsilon_{\pi\eta}$ mixing angle alone.
Regarding the effect of the error from the measured $\pi^{-}\pi^{0}$ VFF onto the $\pi\eta^{(\prime)}$ VFFs,
this is tiny and hence neglected.
The same happens to the scalar contribution,
our values are in accordance with present calculations.
For the total branching ratios, all of them satisfy the current experimental upper bounds.
Finally, in order to test the dependence of our results on the value of the $\varepsilon_{\pi\eta^{(\prime)}}$ mixing angle,
we will also use two sets of different values besides our default ones,
$\varepsilon_{\pi\eta}=(9.8\pm 0.3)\times 10^{-3}$ and $\varepsilon_{\pi\eta^{\prime}}=(2.5\pm 1.5)\times 10^{-4}$.
The first set, named \emph{set 1}, consists of
$\varepsilon_{\pi\eta}=0.0134$ \cite{Paver:2010mz} and $\varepsilon_{\pi\eta^{\prime}}=(3\pm 1)\times 10^{-3}$
\cite{Paver:2011md},
while the second, \emph{set 2}, employs
$\varepsilon_{\pi\eta}=0.0155$ and $\varepsilon_{\pi\eta^{\prime}}=6.79\times 10^{-3}$ \cite{Volkov:2012be}.
Using these sets, we obtain for the vector contribution
BR$_{V}=0.49\times 10^{-5}$ and BR$_{V}=0.66\times 10^{-5}$ for {\emph{set 1}} and {\emph{set 2}}, respectively,
together with new results for the scalar (depending on the SFF employed) and total contributions
which tend to be smaller than, but in agreement with, the reference ones shown in Table~\ref{BRpieta}.

All in all, in view of forthcoming measurements,
our aim is to provide reasonable estimates for the $\tau^{-}\to\pi^-\eta\nu_{\tau}$ branching ratio,
depending on the framework used for the $\pi^-\eta$ SFF,
rather than producing precise results.
Once the $\pi^-\eta$ invariant mass spectrum is available,
it could be used to test the different approaches to this form factor.

\subsection{$\tau^{-}\to\pi^{-}\eta^\prime\nu_{\tau}$}
In Fig.~\ref{distributionpietap}, the total differential decay width distribution of $\tau^{-}\to\pi^{-}\eta^{\prime}\nu_{\tau}$
is shown as a function of the $\pi^-\eta^{\prime}$ invariant mass for several parameterizations of the
$\pi^-\eta^{\prime}$ SFF.
The vector contribution alone is included again.
In this case, the large mass of the $\eta^{\prime}$ considerably reduces the available phase space with respect to 
the $\pi^{-}\eta$ mode.
As a result, the vector contribution is suppressed because the opening of the $\pi^{-}\eta^{\prime}$ production threshold
happens well beyond the region of possible $\rho(770)$ effects. 
In consequence, the $\tau^{-}\to\pi^{-}\eta^\prime\nu_{\tau}$ decay is mainly driven by the SFF.
The scalar contribution as obtained from the Breit-Wigner prescription with two resonances (blue dotted curve)
is in this case small since the $a_{0}(980)$ effects occur before the $\pi^{-}\eta^{\prime}$ threshold
and only the impact of the $a_{0}(1450)$ is noticeable.
The three coupled channels solution (green dot-dashed curve) shows instead a clear peak around $1.4$ GeV 
which vastly dominates the decay.
This effect could be attributed to the $a_{0}(1450)$ resonance as we argued for the $\pi^{-}\eta$ case.
The same behavior is seen for the elastic treatment (solid black curve) though the peak is now less pronounced.

In Table~\ref{BRpietap}, our predictions for the integrated branching ratio of $\tau^{-}\to\pi^{-}\eta^{\prime}\nu_{\tau}$
are given and compared with previous phenomenological analyses and present-day experimental upper limits.
For the vector contribution, we obtain results two orders of magnitude smaller than former calculations
which can be explained by the fact that we are using a value for the $\varepsilon_{\pi\eta^{\prime}}$ mixing angle
one order of magnitude smaller than previous estimates
(see the $\pi\eta$ subsection above for the numerical values employed).
Remember that the normalized version of the $\pi^{-}\eta^{\prime}$ VFF is fixed from data on $\pi^{-}\pi^0$ decays.
Our results for the scalar contribution, which in most cases dominates the total branching ratio,
are in line with existing analyses and fulfill the present limits.

In short, we find that this decay could be of the order of the current experimental upper bound.
We hope that forthcoming experimental information can soon shed light on this mode.
Taking into account the results of our predictions together with the present limits on these
$\tau^{-}\to\pi^{-}\eta^{(\prime)}\nu_{\tau}$ decays,
one could think of discovering them at Belle-II as a first example of measured SCC.

\begin{table}
\centering
\footnotesize
\begin{tabular}{|l|l|l|l|}
\hline
BR$_V$ & BR$_S$ & BR & Reference\\
\hline
$<10^{-7}$ & $[0.2,1.3]\times 10^{-6}$ & $[0.2,1.4]\times 10^{-6}$ & Nussinov, Soffer \cite{Nussinov:2009sn}\\
$[0.14,3.4]\times 10^{-8}$ & $[0.6,1.8]\times 10^{-7}$ & $[0.61,2.1]\times 10^{-7}$ & Paver, Riazuddin
\cite{Paver:2011md}\\
$1.11\times 10^{-8}$ & $2.63\times 10^{-8}$ & $3.74\times 10^{-8}$ & Volkov, Kostunin \cite{Volkov:2012be}\\
\hline\hline
BR$_V$ & BR$_S$ & BR & Our analysis\\
\hline
$[0.3,5.7]\times 10^{-10}$	& $[2\times 10^{-11},7\times 10^{-10}]$	& $[0.5\times 10^{-10},1.3\times 10^{-9}]$
& Breit-Wigner [$a_{0}(980)$]\\
$[0.3,5.7]\times 10^{-10}$	& $[5\times 10^{-11},2\times 10^{-9}]$	& $[0.8\times 10^{-10},2.6\times 10^{-9}]$
& Breit-Wigner  [$a_{0}(980)+a_{0}(1450)$]\\
$[0.3,5.7]\times 10^{-10}$	& $[2\times 10^{-9},4\times 10^{-8}]$	& $[2.6\times 10^{-9},4\times 10^{-8}]$
& Elastic dispersion relation\\
$[0.3,5.7]\times 10^{-10}$	& $[2\times 10^{-7},2\times 10^{-6}]$	& $[2\times 10^{-7},2\times 10^{-6}]$
& 2 coupled channels ($\pi^{-}\eta$ \& $\pi^{-}\eta^{\prime}$)\\
$[0.3,5.7]\times 10^{-10}$	& $[3\times 10^{-7},3\times 10^{-6}]$	& $[3\times 10^{-7},3\times 10^{-6}]$
& 2 coupled channels ($\pi^{-}\eta$ \& $K^{-}K^{0}$)\\
$[0.3,5.7]\times 10^{-10}$	& $[1\times 10^{-7},1\times 10^{-6}]$	& $[1\times 10^{-7},1\times 10^{-6}]$
& 3 coupled channels\\
\hline\hline
& & BR & Experimental collaboration\\
\hline
& & $<4\times 10^{-6}$	($90\%$ CL)	& BaBar \cite{Lees:2012ks}\\
& & $<7.2\times 10^{-6}$ ($90\%$ CL)	& BaBar \cite{Aubert:2008nj}\\
\hline
\end{tabular}
\caption{Vector, scalar and total contributions to the branching ratio (BR) of $\tau^{-}\to\pi^-\eta^{\prime}\nu_{\tau}$.
\emph{Upper part}:
results from previous phenomenological analyses.
\emph{Mid part}:
results from our analysis depending on the description of the $\pi^-\eta^{\prime}$ SFF used.
\emph{Lower part}:
Current experimental upper bounds.}
\label{BRpietap}
\end{table}

\subsection{$\eta^{(\prime)}\to\pi^{+}\ell^{-}\bar{\nu}_{\ell}$ $(\ell=e, \mu)$}
The form factors required for describing $\tau^{-}\to\pi^{-}\eta^{(\prime)}\nu_{\tau}$ decays
and the semileptonic $\eta^{(\prime)}\to\pi^{+}\ell^{-}\bar{\nu}_{\ell}$ $(\ell=e, \mu)$ decays
are the same because the hadronic matrix element $\langle\eta^{(\prime)}|\bar{d}\gamma^{\mu}u|\pi^{+}\rangle$
is related by crossing symmetry with the one in Eq.~(\ref{vectorcurrent}). 
However, in $\eta^{(\prime)}_{\ell3}$ decays the available kinematical energy range is
$m_{\ell}^{2}\le s\le (m_{\eta^{(\prime)}}-m_{\pi})^{2}$ instead of $(m_{\eta^{(\prime)}}+m_{\pi})^{2}\le s\le m_{\tau}^{2}$
for the $\tau$ decays.
Consequently, the form factors entering $\eta_{\ell3}$ decays are by analyticity real functions of $s$.
The differential decay width is given for these decays by
\be
\begin{array}{l}
\displaystyle{\frac{d\Gamma\left(\eta^{(\prime)}\to\pi^{+}\ell^{-}\bar{\nu}_{\ell}\right)}{d\sqrt{s}}=
\frac{G_F^2s^2}{12\pi^3m_{\eta^{(\prime)}}^3}S_{\rm EW}|V_{ud}F_+^{\pi^-\eta^{(\prime)}}(0)|^2
\left(1-\frac{m_\ell^2}{s}\right)^2}\\[2ex]
\qquad
\displaystyle{\times\left[\left(2+
\frac{m_\ell^2}{s}\right)q_{\pi^-\eta^{(\prime)}}^3(s)\widetilde{F}_+^{\pi^-\eta^{(\prime)}}(s)^2+
\frac{m_\ell^2}{s}\frac{3\Delta_{\pi^-\eta^{(\prime)}}^2}{4s}q_{\pi^-\eta^{(\prime)}}(s)
\widetilde{F}_0^{\pi^-\eta^{(\prime)}}(s)^2\right]}\ ,
\end{array}
\label{etal3dist}
\ee
where the VFF contribution highly dominates over the SFF one because this latter is weighted by the squared lepton mass.

\begin{figure}
\centering
\includegraphics[scale=0.4]{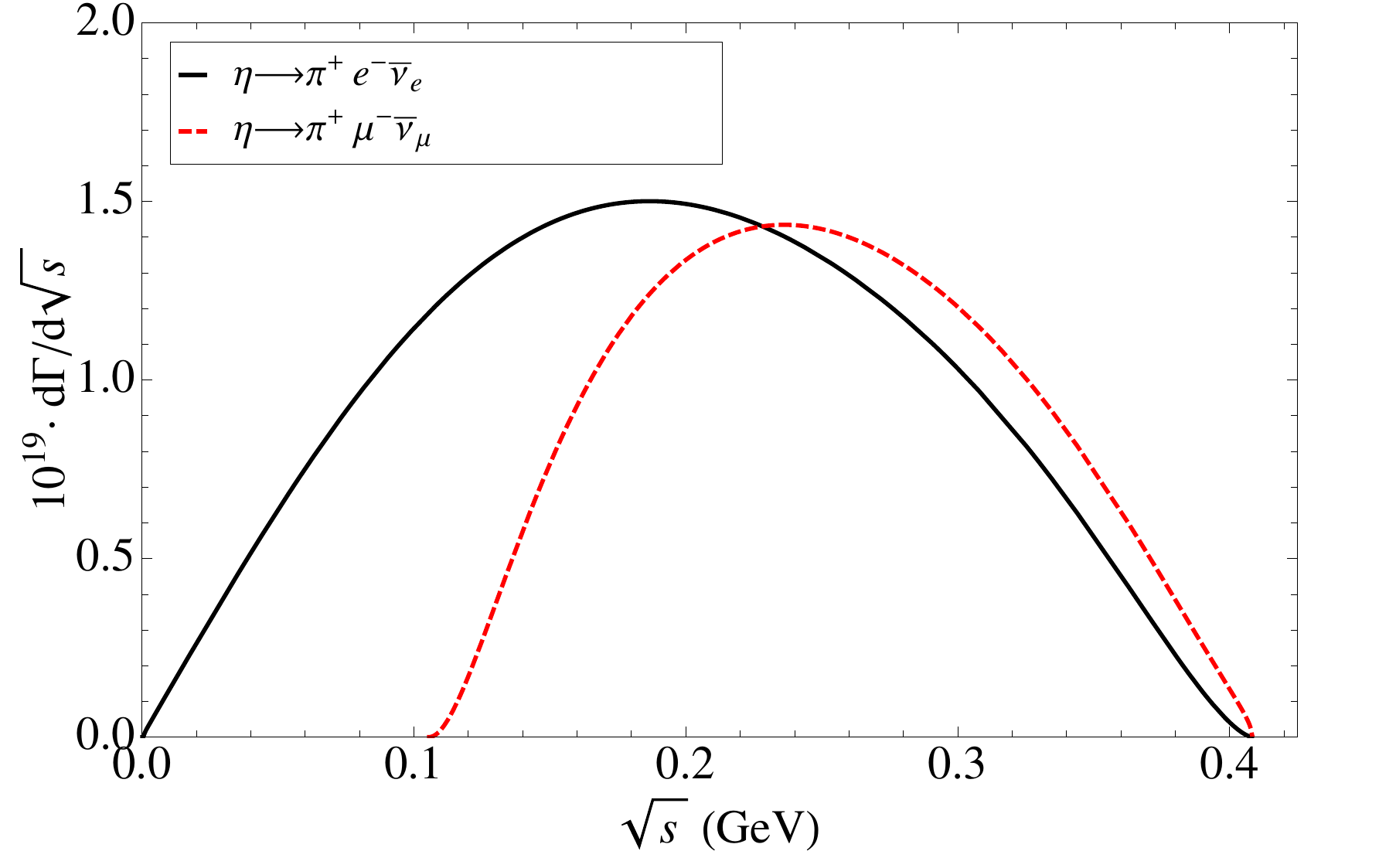}
\includegraphics[scale=0.4]{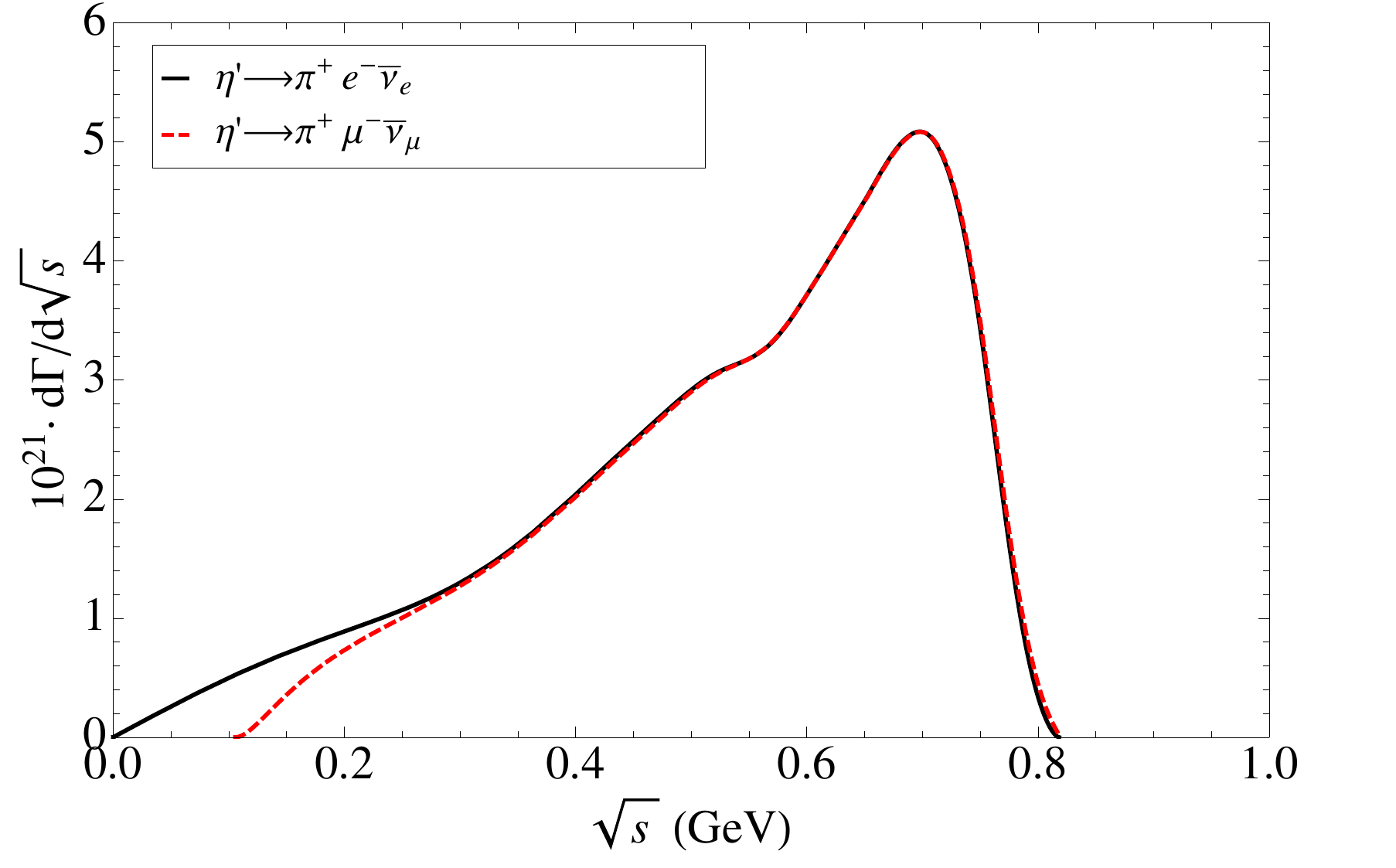}
\caption{Total differential decay width distribution for
$\eta\to\pi^{+}\ell^{-}\bar{\nu}_{\ell}$ (left plot) and $\eta^{\prime}\to\pi^{+}\ell^{-}\bar{\nu}_{\ell}$ (right plot).
The electronic channel (solid black curves) and the muonic one (red dashed curves) are shown.}
\label{etal3dis}
\end{figure}

In Fig.~\ref{etal3dis}, the total differential decay width distributions of
$\eta^{(\prime)}\to\pi^{+}\ell^{-}\bar{\nu}_{\ell}$ $(\ell=e, \mu)$ are displayed.
In Table \ref{etal3predictions}, the results of our analysis for the integrated branching ratios are presented.
These have been obtained after employing
$\varepsilon_{\pi\eta}=(9.8\pm 0.3)\times 10^{-3}$ and $\varepsilon_{\pi\eta^{\prime}}=(2.5\pm 1.5)\times 10^{-4}$
for the $\eta_{\ell3}$ and $\eta^{\prime}_{\ell3}$ channels, respectively.
For the electronic channel, our predictions are compared with the ones obtained in Ref.~\cite{Descotes-Genon:2014tla}.
These are 2.5 times bigger than ours,
which can be easily understood from the fact that the value $\varepsilon_{\pi\eta}=(1.56\pm 0.23)\times 10^{-2}$
is used in this case.
Predictions from Ref.~\cite{Neufeld:1994eg}, with $\varepsilon_{\pi\eta}=1.21\times 10^{-2}$,
lie in the middle\footnote{Due to
the small phase space available, different energy dependences of the normalized $\pi^-\eta^{\prime}$ VFF
do not cause a sizable effect.}.
The rareness of these semileptonic decay modes enhances the sensitivity to new types of interactions
and any clear deviation from branching ratios of order $\mathcal{O}(10^{-13},10^{-12})$
might probe physics beyond the SM.
At the moment, the BESIII Coll.~has reported BR$(\eta\to\pi^{+}e^{-}\bar{\nu}_{e}+{\rm c.c.})<1.7\times 10^{-4}$
and BR$(\eta^{\prime}\to\pi^{+}e^{-}\bar{\nu}_{e}+{\rm c.c.})<2.2\times 10^{-4}$, both at the $90\%$ C.L.,
which are considered as the firsts upper bounds ever for $\eta$ and $\eta^{\prime}$ semileptonic weak decays
\cite{Ablikim:2012vn},
but still extremely far from present estimates.

\begin{table}
\centering
\begin{tabular}{|c|c|c|c|c|}
\hline
Decay & Descotes-Genon, Moussallam \cite{Descotes-Genon:2014tla} & Our analysis\\
\hline
$\eta\to\pi^{+}e^{-}\bar\nu_{e}+{\rm c.c.}$		& $\sim1.40\times 10^{-13}$ & $0.6\times 10^{-13}$\\
$\eta\to\pi^{+}\mu^{-}\bar\nu_{\mu}+{\rm c.c.}$	& $\sim1.02\times 10^{-13}$ & $0.4\times 10^{-13}$\\
\hline
$\eta^{\prime}\to\pi^{+}e^{-}\bar\nu_{e}+{\rm c.c.}$		& & $1.7\times 10^{-17}$\\
$\eta^{\prime}\to\pi^{+}\mu^{-}\bar\nu_{\mu}+{\rm c.c.}$	& & $1.7\times 10^{-17}$\\
\hline
\end{tabular}
\caption{Branching ratio estimates for
$\eta^{(\prime)}\to\pi^{+}\ell^{-}\bar{\nu}_{\ell}$ $(\ell=e,\mu)$ semileptonic weak decays.}
\label{etal3predictions}
\end{table}

\section{Conclusions}
\label{conclusions}
Hadronic decays of the $\tau$ lepton constitute an ideal scenario for studying the hadronization of QCD currents
in its non-perturbative regime.
In this work, we have examined the $\tau^{-}\to\pi^{-}\eta^{(\prime)}\nu_{\tau}$ decays
which, being allowed, though isospin suppressed, SM processes,
belong to the so-called SCC processes unseen in Nature so far.

We have focused on the SM prediction of these decays by describing the participant scalar and vector form factors.
These have been calculated within ChPT including resonances as explicit degrees of freedom
as an initial setup approach.
In this framework, we have encoded the $\pi^{0}$-$\eta$-$\eta^{\prime}$ mixing
by means of three Euler angles $(\varepsilon_{\pi\eta}$, $\varepsilon_{\pi\eta^{\prime}}$ and $\theta_{\eta\eta^\prime})$,
where the $\varepsilon_{\pi\eta^{(\prime)}}$ are isospin-violating quantities,
acting as normalizations of the corresponding form factors, which explain the smallness of these decays.
One interesting consequence which emerges neatly in this parameterization is that the normalized
$\pi^{-}\eta$ and $\pi^{-}\eta^{\prime}$ VFFs are found to be identical to the well-known $\pi^{-}\pi^{0}$ VFF.
Taken advantage of this fact, we have implemented in our study the experimental determination on the latter,
obtained from the Belle collaboration in the analysis of $\tau^{-}\to\pi^{-}\pi^{0}\nu_{\tau}$ decays,
for describing the former in a model-independent way.
Regarding the SFF description, we have discussed different parameterizations according to their increasing
fulfillment of analyticity and unitarity.
We started considering a Breit-Wigner representation by resuming inelastic width effects into the
resonance propagator(s) but neglecting the real part of the corresponding loop function,
hence inducing a violation of both requirements.
This case has been tackled by taking into account, first, the contribution of the $a_{0}(980)$ as the only resonant state
and, second, by including the nearest radial excitation $a_{0}(1450)$ into the representation.
Then, we moved to a completely analytic description, respecting elastic unitarity, by the use of a dispersion relation
through the well-known Omn\`{e}s integral.
This solution requires as an input the elastic phase of the form factor which has been obtained from the
corresponding scattering amplitude after invoking Watson's theorem.
Finally, we have illustrated a method for solving coupled channels form factors by using closed algebraic expressions
after exemplifying the equivalence with the Omn\`es solution for the single channel elastic case.
In this way, the lowest-lying scalar resonances are generated dynamically after considering final-state interactions of
meson-meson systems.

Concerning our predictions for the branching ratios and spectra,
several comments are in order.
For the $\pi^{-}\eta$ decay channel, we have found total BRs of the order of $10^{-5}$,
in agreement with previous theoretical estimates and respecting the current experimental upper bound.
Both vector and scalar contributions are comparable.
While the former is fixed from experiment up to an overall normalization constant,
the $\varepsilon_{\pi\eta}$ mixing angle
which we have computed at next-to-leading order in ChPT including resonances,
and is dominated by $\rho$-exchange,
the value of the latter depends on the SFF description.
The Breit-Wigner approach including one or two scalar resonances give similar results and
these are bigger than the ones obtained from the elastic dispersion relation
(not adding resonances explicitly but generating them dynamically).
We have seen that the effect of coupling the $\pi^{-}\eta^{\prime}$ channel into the $\pi^{-}\eta$ SFF is small
since it does not differ so much from the elastic result.
However, the effect of incorporating the $K^-K^0$ threshold is sizable.
This may be due to the exotic nature of the scalars coupled to the $\bar{u}d$ operator.
For the $\pi^{-}\eta^{\prime}$ decay channel,
this is mainly driven by the scalar contribution because of phase space considerations. 
It is much more sensitive to both the $\varepsilon_{\pi\eta^{\prime}}$ normalization and the SFF description.
We have also seen that inelastic channels may increase the BR of this mode by two orders of magnitude up to
$10^{-6}$.
In any case, accurate predictions of these two processes demand precise values for the
$\varepsilon_{\pi\eta^{(\prime)}}$ mixing angles.
An updated analysis of these two isospin-breaking parameters would be very welcome.
The main drawback of the present work is that the errors associated to the SFFs contributions coming from
the dispersive treatments (elastic or coupled channels)
are underestimated since correlations among the participating parameters are unknown.
This important limitation shall be improved once these decay modes are first measured,
ideally through a joint analysis with the related meson-meson scattering data.

To summarize, considering the tighter bounds on the $\pi^{-}\eta^{(\prime)}$ channels,
both $\tau^-\to\pi^{-}\eta^{(\prime)}\nu_\tau$ decay modes have good prospects for discovering SCC soon at Belle-II.
While the $\rho(770)$ meson shoulder should be an unambiguous signature of this discovery in the $\pi^{-}\eta$ mode,
the thin peak of the $a_0(1450)$ resonance would be very much helpful in both cases. 
Finally, as a by-product, we have also given estimates for the semileptonic crossing symmetric decays
$\eta^{(\prime)}\to\pi^{+}\ell^{-}\bar\nu_{\ell}$ $(\ell=e, \mu)$ for which detection in the near future is not foreseen.
We hope our work will serve as a motivation for the experimental collaborations to measure these decays soon
at Belle-II, BESIII and forthcoming facilities.

\appendix
\section{Form factors in coupled channels analyses}
\label{App}
The once-subtracted dispersion relation for a form factor is written as
\be
F(s)=F(s_{0})+
\frac{s-s_{0}}{\pi}\int_{s_{\rm th}}^{\infty}ds^{\prime}
\frac{{\rm{Im}}F(s^{\prime})}{(s^{\prime}-s_{0})(s^{\prime}-s-i\epsilon)}\ ,
\label{CC1}
\ee
where $F(s)$ is now, in the case of coupled channels, a $n$-entries column vector.
Besides analyticity, the form factor can also satisfy unitarity.
The unitarity relation ${\rm{Im}}F(s)=\Sigma(s)t_{IJ}^{\ast}(s)F(s)$,
with $\Sigma(s)$ a diagonal matrix of kinematical factors given by
\be
\Sigma(s)=
\begin{pmatrix}
             \sigma_{1}(s) & 0 &\cdots & 0\\[1ex]
             0 & \sigma_{2}(s) &\cdots & 0\\[1ex]
             \cdots &\cdots &\cdots & 0\\[1ex]
             0 & 0 & 0 &\sigma_{n}(s)
\end{pmatrix}\ ,
\ee
and $t_{IJ}(s)$ a $n\times n$ matrix defined as
\be
t_{IJ}(s)=
\begin{pmatrix}
             t^{11}(s) & t^{12}(s) &\cdots & t^{1n}(s)\\[1ex]
             t^{21}(s) & t^{22}(s) &\cdots & t^{2n}(s)\\[1ex]
             \cdots &\cdots &\cdots &\cdots\\[1ex]
             t^{n1}(s) & t^{n2}(s) &\cdots & t^{nn}(s)
\end{pmatrix}\ ,
\ee
encoding the required unitarised partial-wave amplitudes,
allows to rewrite the form factor as
\be
F(s+i\epsilon)=F(s_{0})+
\frac{s-s_{0}}{\pi}\int_{s_{\rm th}}^{\infty}ds^{\prime}
\frac{\Sigma(s^{\prime})t_{IJ}^{\ast}(s^{\prime})F(s^{\prime})}{(s^{\prime}-s_{0})(s^{\prime}-s-i\epsilon)}\equiv
F(s_{0})+\widetilde{F}(s+i\epsilon)\ ,
\label{CC2}
\ee
where $F(s_{0})$ is real and the discontinuity of $\widetilde{F}(s+i\epsilon)$ is given by
\be
\widetilde{F}(s+i\epsilon)-\widetilde{F}(s-i\epsilon)=
2i\lim_{\epsilon\to 0}{\rm{Im}}F(s+i\epsilon)=2i{\rm{Im}}F(s)=2i\Sigma(s)t_{IJ}^{\ast}(s)F(s)\ .
\label{discontinuity}
\ee
To unitarize the partial-wave scattering amplitudes, we introduce the $N/D$ method
\be
t_{IJ}(s)=\frac{N_{IJ}(s)}{D_{IJ}(s)}\ ,
\label{ND}
\ee
where the matrix functions (the $IJ$ indices are omitted hereafter)
$N$ and $D$ contain the left- and right-hand cuts of the partial-wave amplitudes, respectively,
and satisfy the dispersion relations
\be
N(s)=\frac{s-s_{0}}{\pi}\int_{-\infty}^{s_{L}}ds^{\prime}
\frac{{\rm{Im}}N(s^{\prime})}{(s^{\prime}-s_{0})(s^{\prime}-s-i\epsilon)}
\ee
and
\be
D(s)=D(s_{0})+\frac{s-s_{0}}{\pi}\int_{s_{\rm th}}^{\infty}ds^{\prime}
\frac{{\rm{Im}}D(s^{\prime})}{(s^{\prime}-s_{0})(s^{\prime}-s-i\epsilon)}\ .
\label{disD}
\ee
Unitarity applied to the inverse of the partial-wave amplitudes fulfills ${\rm{Im}}t^{-1}(s)=-\Sigma(s)$,
or, equivalently,
\be
{\rm{Im}}D(s)=-N(s)\Sigma(s)\ .
\label{ImD}
\ee
By inserting Eq.~(\ref{ImD}) into Eq.~(\ref{ND}), one gets
\be
t^{\ast}(s)=\frac{N^{\ast}(s)}{D^{\ast}(s)}=
\frac{-({\rm{Im}}D(s)/\Sigma(s))^{\ast}}{D^{\ast}(s)}=\frac{-{\rm{Im}}D(s)/\Sigma(s)}{D(s-i\epsilon)}\ .
\label{tTrans}
\ee
Then, using Eq.~(\ref{tTrans}), one rewrites Eq.~(\ref{discontinuity}) as
\be
\widetilde{F}(s+i\epsilon)-\widetilde{F}(s-i\epsilon)=
2i\Sigma(s)\left[\frac{-{\rm{Im}}D(s)/\Sigma(s)}{D(s-i\epsilon)}\right]\left[F(s_{0})+\widetilde{F}(s+i\epsilon)\right]\ ,
\label{discontinuity2}
\ee
which further reduces to
\be
\widetilde{F}(s+i\epsilon)\left[D(s-i\epsilon)+2i{\rm{Im}}D(s)\right]-
\widetilde{F}(s-i\epsilon)D(s-i\epsilon)=-2i{\rm{Im}}D(s)F(s_{0})\ .
\label{termbracket}
\ee
Once the term in square brackets is written as $D(s+i\epsilon)$ (the discontinuity across the cut),
one arrives at the following expression
\be
\widetilde{F}(s+i\epsilon)D(s+i\epsilon)-\widetilde{F}(s-i\epsilon)D(s-i\epsilon)=
-2i{\rm{Im}}D(s)F(s_{0})\ ,
\ee
whose once subtracted solution, as a result of the Cauchy integral, reads
\be
\widetilde{F}(s+i\epsilon)D(s+i\epsilon)=
\frac{s-s_{0}}{2\pi i}\int_{s_{\rm th}}^{\infty}ds^{\prime}
\frac{F(s^{\prime}+i\epsilon)D(s^{\prime}+i\epsilon)-F(s^{\prime}-i\epsilon)D(s^{\prime}-i\epsilon)}
{(s^{\prime}-s_{0})(s^{\prime}-s)}\ ,
\ee
and hence
\be
\widetilde{F}(s+i\epsilon)=\frac{1}{D(s+i\epsilon)}\frac{-(s-s_{0})}{\pi}\int_{s_{\rm th}}^{\infty}ds^{\prime}
\frac{{\rm{Im}}D(s^{\prime})F(s_{0})}{(s^{\prime}-s_{0})(s^{\prime}-s)}\ ,
\ee
which employing Eq.~(\ref{disD})  brings to
\be
\widetilde{F}(s+i\epsilon)=-D(s+i\epsilon)^{-1}\left[D(s+i\epsilon)-D(s_{0})\right]F(s_{0})\ .
\ee
Finally, the form factor in Eq.~(\ref{CC2}), after imposing analyticity and unitarity,
is found to be
\be
F(s)=F(s_{0})-D(s)^{-1}\left[D(s)-D(s_{0})\right]F(s_{0})=D(s)^{-1}D(s_{0})F(s_{0})\ .
\label{coupledchannels}
\ee
As written in Eq.~(\ref{coupledchannels}),
the form factor's problem in a coupled channels analysis is reduced to finding a suitable parameterization for the
$D(s)$ matrix.
In this work, we have used
\be
D_{IJ}(s)=\mathbbm{1}+g(s)N_{IJ}(s)\ ,
\label{DIJmatrix}
\ee
in analogy with Eq.~(\ref{elasticclosed}) for the single channel case.
The matrices $g(s)$ and $N_{IJ}(s)$ are generalizations for the multichannel case of the definitions given there.

\section*{Acknowledgements}
This work was supported in part by the FPI scholarship BES-2012-055371 (S.G-S),
the Secretaria d'Universitats i Recerca del Departament d'Economia i Coneixement de la Generalitat de Catalunya
under grant 2014 SGR 1450,
the Ministerio de Ciencia e Innovaci\'on under grant FPA2011-25948,
the Ministerio de Econom\'{i}a y Competitividad under grants CICYT-FEDER-FPA 2014-55613-P and SEV-2012-0234,
the Spanish Consolider-Ingenio 2010 Program CPAN (CSD2007-00042),
and the European Commission under program FP7-INFRASTRUCTURES-2011-1 (Grant Agreement N. 283286).
P.~R.~acknowledges financial support from
projects 236394 (Conacyt, Mexico), 250628 (Ciencia B\'{a}sica) and 296 (Fronteras de la Ciencia)
as well as the hospitality of IFAE, where part of this work was done.

\end{document}